\documentclass[fleqn,usenatbib]{mnras}

\usepackage{newtxtext,newtxmath}

\usepackage[T1]{fontenc}
\DeclareRobustCommand{\VAN}[3]{#2}
\let\VANthebibliography\thebibliography
\def\thebibliography{\DeclareRobustCommand{\VAN}[3]{##3}\VANthebibliography}

\usepackage{graphicx}	
\usepackage{amsmath}	

\makeatletter
\newcommand\thickhline{%
  \noalign{\ifnum0=`}\fi\hrule \@height \thickarrayrulewidth \futurelet
   \reserved@a\@xthickhline}
\def\@xthickhline{\ifx\reserved@a\thickhline
               \vskip\doublerulesep
               \vskip-\thickarrayrulewidth
             \fi
      \ifnum0=`{\fi}}
\makeatother
\newlength{\thickarrayrulewidth}
\usepackage{ulem, orcidlink}

\title[NGC\,5253: nitrogen enrichment at pc scales]{The heart of NGC\,5253 as seen with MUSE-NFM: nitrogen enrichment through stellar chemical feedback at parsec scales}

\author[B.G.A. Pruijt et al.]{Brigitte G.A. Pruijt\orcidlink{0009-0007-5947-2343},$^{1}$\thanks{E-mail: pruijt@strw.leidenuniv.nl}
Ana Monreal-Ibero\orcidlink{0000-0002-6455-2491},$^{1}$
Peter M. Weilbacher,$^{2}$
Jeremy R. Walsh\orcidlink{0000-0002-8008-910X},$^{3}$
Sebastian Kamann\orcidlink{0000-0001-6604-0505},$^{4}$
\newauthor
Azlizan A. Soemitro\orcidlink{0000-0001-6213-4117},$^{2,5}$
Haruka Kusakabe\orcidlink{0000-0002-3801-434X},$^{6}$
Leindert Boogaard\orcidlink{0000-0002-3952-8588}$^{1}$
\\
$^{1}$Leiden Observatory, Leiden University, P.O. Box 9513, 2300 RA Leiden, The Netherlands\\
$^{2}$Leibniz-Institut für Astrophysik Potsdam (AIP), An der Sternwarte 16, 14482 Potsdam, Germany\\
$^{3}$European Southern Observatory, Karl-Schwarzschild Strasse 2, 85748 Garching, Germany\\
$^{4}$Astrophysics Research Institute, Liverpool John Moores University, IC2 Liverpool Science Park, 146 Brownlow Hill, Liverpool L3 5RF, UK\\
$^{5}$Institut für Physik und Astronomie, Universität Potsdam, Karl-Liebknecht-Str. 24/25, 14476 Potsdam, Germany\\
$^{6}$Department of General Systems Studies, Graduate School of Arts and Sciences, The University of Tokyo, 3-8-1 Komaba, Meguro-ku, Tokyo, 153-8902, Japan
}

\date{Accepted XXX. Received YYY; in original form ZZZ}

\pubyear{2025}

\begin{document}
\label{firstpage}
\pagerange{\pageref{firstpage}--\pageref{lastpage}}
\maketitle

\begin{abstract}
NGC\,5253 is a nearby (D=3.6\,Mpc) Blue Compact Dwarf galaxy, notable for its three massive young super star clusters (SSCs) and nitrogen enrichment. Its similarity to extreme star-forming galaxies at high redshift makes it a good local analogue for studying chemical enrichment at high spatial resolution. We characterise the ionised gas and dust in the giant \ion{H}{II} region in the proximity of the three SSCs in the centre of NGC\,5253 using new Multi-Unit Spectroscopic Explorer Narrow Field Mode adaptive optics-assisted data at unprecedented spatial resolution of 0\farcs15$\sim$2.3\,pc. We derive the attenuation for the central SSCs and, for the first time, map the extinction parameter ($R_V$) in an extragalactic object. $R_V$ varies among SSCs, suggesting differences in dust physics. Electron temperature and density diagnostics yield flat temperature distributions $T_\mathrm{e,median}$([\ion{N}{II}])$=12000 \pm 1700$\,K and $T_\mathrm{e,median}$([\ion{S}{III}])$ = 11000 \pm 600$\,K, and a structured $n_e$([\ion{S}{II}]) of maximum $1930 \pm 40$\,cm$^{-3}$. The direct method gives a flat helium abundance ($10^3y^+ = 81 \pm 4$) and uniform oxygen abundance ($12 + \log(\text{O/H}) = 8.22 \pm 0.05$). N/O shows a factor 2–3 enhancement around the SSCs, mapped here for the first time at such high spatial resolution. The total excess nitrogen mass is $\sim$0.3\,$M_\odot$, which we estimate is producible by the observed WN-type Wolf–Rayet (WR) stars. Because there is no direct spatial overlap between the enrichment and WR star positions, the N-rich material appears to have been expelled from the original sites.
\end{abstract}





\section{Introduction}
Blue Compact Dwarf (BCD) galaxies are dwarf ($M_\star \lesssim 10^8\,M_\odot$) galaxies currently undergoing an episode of compact star formation (starburst diameter $\lesssim$1 kpc). They typically have low oxygen abundances of $12+\log(O/H)\lesssim 7.9$, with values as low as $12+\log(O/H)\approx 7.0$ \citep[e.g.][]{izotov_j08114730_2018}. Furthermore, they are defined by a limiting magnitude of $M_B \approx -18$ and exhibit strong emission lines similar to those found in \ion{H}{II} regions. These are all properties that BCD galaxies share with the very first galaxies in the early Universe \citep{izotov_low-redshift_2021}. Since BCD galaxies are found in the local Universe, they provide a unique opportunity to explore the interplay between gas, dust, and star formation at a level of detail that is unattainable at high redshift. Specifically, they are ideal reference targets for studying stellar chemical, mechanical, and radiative star formation (SF) feedback processes at high redshift. 

Detailed studies of the chemical composition of galaxies are crucial for our understanding of galaxy evolution. Imprints left by chemical feedback in galaxies, in the form of abundance ratios, trace the products of nucleosynthesis in stars. Of particular interest are observations of carbon, oxygen, and nitrogen, which are among the most abundant elements in the universe. The nitrogen-to-oxygen ratio (N/O) is shown to depend on the gas phase metallicity (O/H), through observations and numerical modelling (e.g. \citealt{izotov_chemical_2006, pilyugin_counterpart_2012, vincenzo_nitrogen_2016}). For low metallicities, the relation between N/O and O/H is flat, but at metallicities of 12+log(O/H)$\gtrapprox8$ evolves into a positive relation. This trend arises from the production of nitrogen via two main processes: a primary contribution from massive stars and a secondary contribution from asymptotic giant branch (AGB) stars (e.g. \citealt{vincenzo_nitrogen_2016}). 

With the advent of the James Webb Space Telescope (JWST), chemical abundances can now be observed up to very high redshifts (z$\gtrsim$4). In these galaxies, studies show values of log(N/O) that are significantly higher than expected for the metallicity of the galaxy (e.g. \citealt{bunker_jades_2023,marques-chaves_extreme_2024,ji_ga-nifs_2024,schaerer_discovery_2024,2024watanabe}). The most notable example is GN-z11 at $z\approx10.6$ with log(N/O)$\gtrapprox$-0.5 (\citealt{cameron_nitrogen_2023,senchyna_gn-z11_2024}). The origin of this nitrogen excess is still debated, with various hypotheses being investigated. These include the ejection of CNO-processed gas via stellar winds in massive, super massive, and Wolf-Rayet (WR) stars (e.g. \citealt{izotov_chemical_2006}), a connection to proto-globular clusters (GCs; e.g. \citealt{senchyna_gn-z11_2024}), broad-line AGN \citep{isobe_jades_2025}, luminous blue variable stars, and even tidal disruption events \citep{cameron_nitrogen_2023}. 

Despite major technological advancements, it remains impossible to study the nitrogen enrichment at scales smaller than galaxy size at high redshift. However, many questions remain. How does the chemical enrichment vary on small scales? Can it be tied to the presence of WR or (super) massive stars? What happens in regions of extreme SF? To answer these questions and understand the origin of nitrogen overabundance in primeval galaxies, resolved studies are necessary. For this, we can leverage the analogy of primeval galaxies with BCD galaxies. 

Perhaps one of the most famous example of a BCD galaxy showing nitrogen enrichment is NGC\,5253 \citep[e.g.][]{campbell_stellar_1986, walsh_optical_1989,kobulnicky_hubble_1997, lopezsanchez_localized_2007, monreal-ibero_study_2010}. In this study, we explore the chemical feedback in this galaxy at unprecedented spatial resolution using data from Multi Unit Spectroscopic Explorer (MUSE), Narrow Field Mode, taken with adaptive optics. 

At a distance of just $3.6 \pm 0.2$ Mpc \citep{sakai_effect_2004}, NGC\,5253 is the closest example of a dwarf star-forming galaxy displaying nitrogen enrichment. The galaxy is located in the Cen\,A/M83 galaxy complex \citep{karachentsev_hubble_2007}. NGC\,5253 is currently undergoing a burst of star formation, likely triggered by a past interaction, as evidenced by the \ion{H}{I} streamer system in the halo of which a CO streamer appears \citep{kobulnicky_inflows_2008,lopez-sanchez_intriguing_2012}. The basic characteristics of NGC\,5253 are shown in Table \ref{tab:properties_NGC5253}. Given its mass and metallicity, NGC\,5253 falls on the large and metal-rich end of the BCD galaxies. It is well studied, with a wealth of multi-wavelength ancillary data available that covers the complete spectrum. 

The central region of NGC\,5253 hosts a large number of relatively young stellar clusters, with ages $<$5\,Myr \citep{harris_recent_2004, de_grijs_ngc_2013}. The starburst itself is confined to a compact 60\,pc region \citep{calzetti_dust_1997, calzetti_structure_1999, tremonti_star_2001}. In addition, a small number of older (globular) clusters, aged $\sim 1$\,Gyr and $\sim10$\,Gyr, have been found at larger radii \citep[][]{harbeck_intermediate_2012, de_grijs_ngc_2013}. This indicates that NGC\,5253 is an old galaxy that has undergone multiple episodes of star formation in its past. 

\begin{figure}
    \centering
    \includegraphics[width=0.45\textwidth]{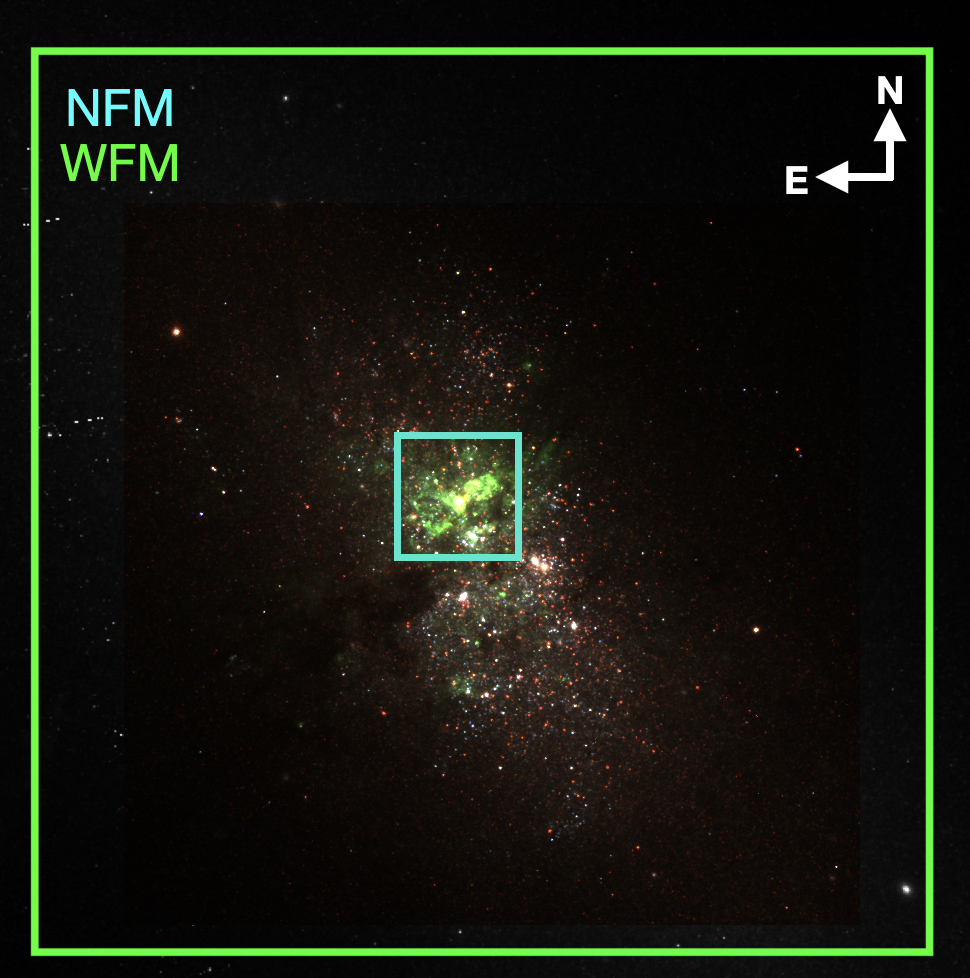}
    \caption{HST ACS Wide Field Camera (WFC) and High Resolution Camera (HRC) RGB image of NGC\,5253. Colours are red: F814W (HRC) / green: F555W (WFC) / blue: F435W (HRC). The cyan and green squared indicate the field of view of the MUSE Narrow Field Mode and Wide Field Mode respectively (see Section \ref{sec:obs}).}
    \label{fig:HST}
\end{figure}

Of particular interest is the central star-forming region, highlighted by the cyan rectangle in Fig. \ref{fig:HST}. This region is crossed by a dust lane towards the South-East. At larger radii, an ionisation cone has been detected \citep{zastrow_ionization_2011}, as well as infalling molecular gas \citep{meier_molecular_2002, miura_witness_2015}.

The central region contains three super star clusters (SSCs), with ages around 1\,Myr. \cite{calzetti_brightest_2015} conducted a detailed study of the two optically visible clusters, designated as cluster \#5 and cluster \#11. Cluster \#5 corresponds to the peak in H$\alpha$ emission in the HST image, while cluster \#11 aligns with the peak in Paschen (Pa) $\alpha$ emission. The third SSC is obscured at optical wavelengths but can and has been studied at radio wavelengths \citep{turner_radio_2000,consiglio_alma_2017} using the Very Large Array (VLA) and the Atacama Large Millimeter/Submillimeter Array (ALMA). This supernebula is the youngest of the three SSCs, with an age $<$1\,Myr and still embedded in its formation dust cloud and is a potential proto-globular cluster. 

Due to uncertainties in astrometry and the necessity to observe the clusters using different instruments, it has long been debated whether there are two or three clusters. One of clusters \#5 or \#11 has often been associated with the supernebula. \cite{smith_three_2020} used the Gaia Data Release 2 catalogue to show that there are three distinct SSCs. In this paper, we follow the nomenclature by \cite{calzetti_brightest_2015} and refer to the SSCs by the names cluster \#5 (the eastern cluster, best visible at optical wavelengths), the supernebula (the SSC best visible in radio that is located to the west of cluster \#5), and cluster \#11 (the most western of the clusters and best visible at IR wavelengths). Because this galaxy is widely studied, the same region in the galaxy can be referred to by many names. To avoid confusion, Table \ref{tab:cluster_names} provides an overview of the names of the SSCs as used in various papers. 

The central region of NGC\,5253 is also the area where the factor 2-3 nitrogen enhancement is observed. Surrounding this area of high N/O, there are reports of a population of WR stars \citep{schaerer_detection_1997,monreal-ibero_study_2010, westmoquette_piecing_2013}, suggesting that they are the origin of the nitrogen enhancement. However, \cite{monreal-ibero_study_2010} (hereafter MI10) show that not all WR stars seem to be the cause of nitrogen enrichment due to a mismatch in spatial distribution. A more detailed study, in particular at high spatial resolution, of this area is required to disentangle the true interplay between the ISM and the stars.

In this paper, we investigate the heart of NGC\,5253 at high spatial resolution using MUSE data. The details of the data acquisition and reduction are described in Sect. \ref{sec:obs}. In Sect. \ref{sec:results} we describe the results of our analysis, specifically, the attenuation (Sect. \ref{sec:results_attenuation}), electron temperature and density (Sect. \ref{sec:results_Te_ne}), and abundance determination (Sect. \ref{sec:results_abundances}). Sect. \ref{sec:discussion_attenuation} discusses the geometry and physical properties of the dust, and Sect. \ref{sec:discussion_Te} focusses on temperature-temperature relations. The origin of nitrogen enhancement will be addressed in Sect. \ref{sec:discussion_NO}. Finally, we present our conclusions in Sect. \ref{sec:conclusions}.

\begin{table}
    \centering
    \caption{Summary of properties of NGC\,5253. (a) NASA/IPAC Extragalactic Database (NED); (b) \citealt{sakai_effect_2004}; (c) \citealt{lopez-sanchez_intriguing_2012}; (d) We assumed 12+log(O/H)$_\odot$=8.66 \citep{asplund_line_2004}; (e) \citealt{marasco_shaken_2023}; $^*$ corrected for the assumed distance. }
    \begin{tabular}{c c c }
    \thickhline
        Parameter & Value & Reference \\
    \thickhline
        Name & NGC\,5253 & (a)\\
        RA (J2000.0) & 13h39m55.9s & (a) \\
        Dec (J2000.0) & -31d38m24s & (a) \\
        $z$ & 0.001358 & (a) \\
        D (Mpc) & 3.6 $\pm$ 0.2 & (b) \\
        scale (pc/") & 17.5 & \\
        12 + log(O/H) & 8.30 $\pm$ 0.02 & (c)\\
        $Z/Z_\odot$ & 0.4 & (c),(d) \\
        log(M$_\star$/$M_\odot$) & $8.66\pm0.24$ & (e)$^*$\\ 
        log(SFR/$M_\odot yr^{-1}$) & $-0.24\pm0.15$ & (e)$^*$\\
    \thickhline
    \end{tabular}
    \label{tab:properties_NGC5253}
\end{table}

\begin{table}
    \centering
    \caption{Nomenclature of the SSCs in other papers.
    C97: \citealt{calzetti_dust_1997}; K97: \citealt{kobulnicky_hubble_1997}; H04: \citealt{harris_recent_2004}; M10: \citealt{monreal-ibero_study_2010}; G13: \citealt{de_grijs_ngc_2013}; C15: \citealt{calzetti_brightest_2015}; AH04: \citealt{alonsoherrero_obscured_2004}.}
    \begin{tabular}{c c c c c c c c}
    \thickhline
       Cluster & K97 & C97 & H04 & M10 & G13 & C15 & AH04\\
    \thickhline
        \#5 & - & 5 & 1 & 1 & 87 & 5 & C1\\
        \#11 & - & - & 1 & 1 & - & 11 & C2\\
    \thickhline
    \end{tabular}
    \label{tab:cluster_names}
\end{table}

\section{Observations and data processing} \label{sec:obs}
\subsection{Data}\label{sec:resolution}
The centre of NGC\,5253 was observed with MUSE \citep{Bacon2010} in narrow-field mode (NFM) on April 21st, 2023. Four exposures of 507\,s were taken during conditions of thin cirrus in external seeing around 0\farcs85 at airmass below 1.015, they were rotated by 90\degr{} after each exposure. An offset sky field was exposed for 54\,s at the end of the observing block. The standard star GD\,71 was observed for flux calibration in the morning after the science exposures during clear conditions. MUSE covers a wavelength range $4750 < \lambda < 9350$\,\AA{}, with a gap between 5870 and 6050\,\AA{} to block out the laser light used for adaptive optics. The sampling is about 25\,mas in the spatial direction and 1.25\,\AA{} in wavelength, each exposure covers about $7\farcs5\times7\farcs5$ on the sky.

In NFM the atmospheric turbulence is modelled using a laser-tomographic adaptive optics system \citep{Stuik2006} which typically delivers spatial resolution of $<0\farcs1$ in the central core of the PSF and Strehl ratios of up to 10\% \citep{wevers_musenfm_2022}.

Data reduction followed standard procedures for MUSE data. We used the pipeline \citep[v2.8.5,][]{weilbacher_musepipeline_2020} for all calibrations, starting from the raw data, using a home-grown script to tie the steps together in a semi-automatic way. The steps included bias subtraction, flat-fielding (through internal lamp and twilight sky), wavelength calibration, geometric positioning of the optical elements, and computation of the line-spread function, with default parameters and using calibrations closest in time to the science data. Each observed sky exposure was then combined for all 24 subfields, without removing the incomplete wavelengths. Atmospheric refraction was not corrected in software, since the correction by the optical system was judged to be good enough. The exposures were then flux calibrated and corrected for telluric absorption. The offset sky field was used for an initial fit to the sky-emission lines (excluding lines below 5197\,\AA) and to derive a sky continuum, for $4600-9370$\,\AA. The sky lines were adapted to each science exposure, using the darkest 20\% of the field to create a sky spectrum to re-fit. No attempts were made to model the telluric Raman lines created by the lasers of the AO system \citep{vogt_raman_2017}. They are hence subtracted as a constant across the field data as part of the sky continuum. The data was then corrected to barycentric velocities, astrometrically calibrated to remove distortions and then resampled into separate cubes. Images reconstructed in the F814W filter from the individual cubes were then aligned, using a Moffat-fit to a star in the MUSE images, with the corresponding HST ACS image as a reference. With these alignment offsets known, all exposures were then resampled from the intermediate pixel tables into one common datacube, with a final extent of $\sim\,8\farcs1\times7\farcs5$\footnote{Due to the rotational dither pattern, the cube is a bit larger than quoted, the edges do not contain data.} with a final wavelength range of $4750-9300$\,\AA{}. The astrometry agrees with Gaia DR3 \citep{gaia_dr3_2023} to within about 35\,mas as judged from over-plotting the catalogue sources on the MUSE white-light image.

We determine the spatial resolution of the data by performing a 2D Moffat fit to point sources in the data. We find a FWHM ranging between 0\farcs22 (for 5000\,\AA) and 0\farcs10 (for 9000\,\AA), with a median resolution of 0\farcs15, corresponding to physical scales of 2.3\,pc at the assumed distance of NGC\,5253. This is enough to resolve individual star clusters. We confirm that the contribution from the wings of the PSF is negligible compared to the background nebular emission. The resolution measurements are consistent between fitting with a Moffat or Maoppy \citep{fetick_physics-based_2019} profile. 

In addition to the NFM, we utilise the wide-field mode (WFM) configuration observations of NGC\,5253 from the archive. It was observed on August 20th, 2015 (PI: Vanzi). Nine exposures were taken, with a total exposure time of 1620\,s. The seeing was 1\farcs0. Contrary to the NFM, the WFM is not assisted by adaptive optics, because of this the data covers the full wavelength range $4750 < \lambda < 9350$\,\AA{}. The data were processed using the standard ESO MUSE pipeline \citep{weilbacher_musepipeline_2020}, and downloaded directly as processed datacubes from the ESO Science Portal. 

Finally, we correct both cubes for the foreground Galactic extinction using $A_V=0.153$ \citep{schlafly_measuring_2011} and $R_V=3.1$ and the \cite{fitzpatrick_correcting_1999} extinction law.

\subsection{Post-reduction data processing} 
To optimise the signal-to-noise ratio (S/N) of the spectra, we perform Voronoi tessellation on the data, utilising the \texttt{Vorbin} module \citep{cappellari_adaptive_2003}. This algorithm partitions the data such that each bin meets a specified minimum S/N for a given spectral feature. In our case, we tessellate our data into bins with a S/N of H$\alpha$ of 600. Tiles encompass a few to a few hundred pixels.
For our analysis, we are interested in the emission lines that trace the ionised gas. The observed spectra comprise multiple components: the stellar (and nebular) continuum, stellar absorption features (most significant for H$\alpha$ and H$\beta$), emission lines from ionised gas, and additional atomic or molecular absorption features. To isolate the emission lines, it is necessary to disentangle these components. To this aim, we use the population spectral synthesis (PSS) code \texttt{FADO}\footnote{\url{https://spectralsynthesis.org/fado.html}} (Fitting Analysis using Differential evolution Optimization; \citealt{gomes_fitting_2017}). One advantage of FADO over other available PSS codes is that FADO includes the contribution of the nebular continuum -- which can contribute on the order of one-third to the optical continuum -- to its spectral fitting; this ensures consistency between the best-fitting stellar model and the observed nebular emission. This is particularly important for galaxies with intense SF activity, such as BCD galaxies. 

\texttt{FADO} allows user-defined modifications to its default input and output parameters. We adjust several parameters, as detailed below. For the base spectra that we put in, we decide to restrict to those with a metallicity $0.008<Z<0.004$, reflecting the low metallicity of NGC\,5253. We used a set of single star populations provided by \cite{bruzual_stellar_2003}, based on Padova 2000 evolutionary tracks \citep{girardi_evolutionary_2000}, while assuming a Salpeter initial mass function between 0.1 and 100\,$M_\odot$ \citep{Salpeter1955}. This set of requirements results in a total of 50 base spectra, with ages ranging from 1 Myr to 15 Gyr. For the spectral fitting, we adopt the wavelength range $4750\,\AA < \lambda < 9300\,\AA$. However, we exclude the interval $5781\,\AA < \lambda < 6048\,\AA$, which is masked due to contamination by the sodium laser guide star used for adaptive optics. Additionally, we change the normalisation wavelength to 5520\,\AA, as \texttt{FADO}’s default of 4020\,\AA\, lies outside the MUSE spectral range.

The resulting best-fit spectra of the stellar continuum are resampled to match the original MUSE spectral sampling and subsequently subtracted. The resulting spectra are used to perform the analysis described in the rest of this paper.

\subsection{Fitting emission lines}
We determined the properties of the emission lines from the continuum-subtracted spectra. To do this, we fit the lines with a single Gaussian using the \texttt{Python} package \texttt{LMFIT}\footnote{\url{https://lmfit.github.io/lmfit-py/intro.html}}. 
Where possible, lines were fitted independently and non-simultaneously and the underlying continuum was modelled as a one-degree polynomial. However, when there was only a small separation in wavelength between emission lines, the lines were fitted simultaneously. Furthermore, all doublets of the same ion were fitted simultaneously and imposed to have the same width. For [\ion{N}{II}]$\lambda$5755, we deviate from the simple one-degree polynomial fit for the continuum and instead fit a spline with 4 knots, to account for the Wolf-Rayet bump. Finally, the [\ion{O}{I}]$\lambda6300,\lambda6364$ sky lines have not been subtracted properly, negatively influencing the fit. Therefore, we replace the continuum at this positions by the median value of the continuum in this part of the spectrum.

All fits were weighed by the statistical error on the spectra as delivered by the MUSE pipeline. We fit for (i) the central wavelength (that is translated to velocities); (ii) the flux; (iii) line widths. For the error on the fit, we use the one returned by \texttt{LMFIT}, based on the covariance matrix. This is true for all properties of the fit. We define the \textit{signal-to-noise (S/N)} of a line as the ratio of the flux over the standard deviation of the continuum close to that line. 

In a select few locations, we used a double Gaussian to fit H$\alpha$ and H$\beta$ and \ion{He}{I}$\lambda$6678, where there is a strong and broad secondary component. These locations correspond to cluster \#84 and \#106 as detected by \citet[][hereafter dG13]{de_grijs_ngc_2013}.

    \section{Results} \label{sec:results}
    \subsection{Overall morphology}
\begin{figure*}
    \centering
    \includegraphics[width=\textwidth]{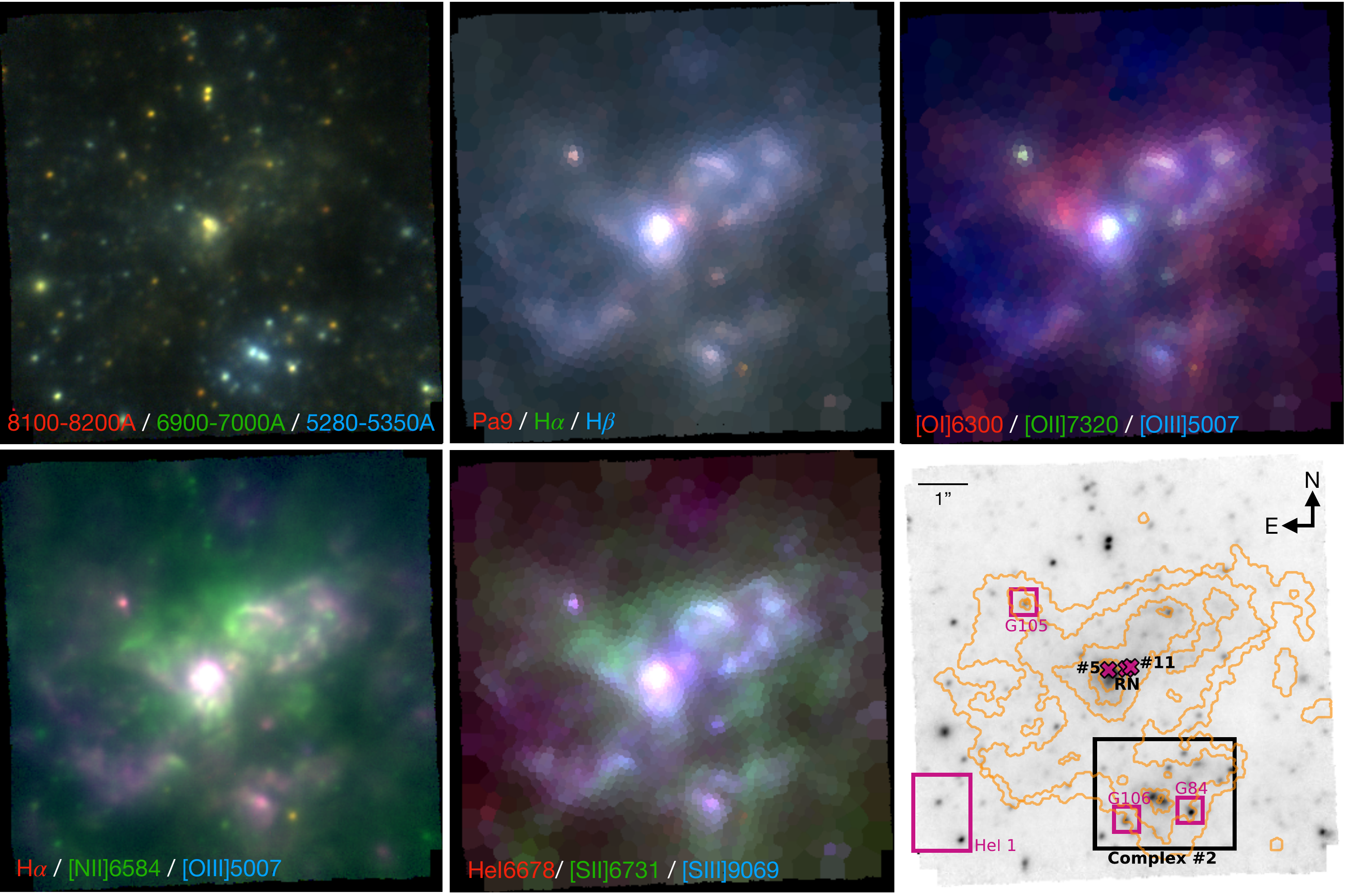}
    \caption{A set of composite images of NGC\,5253. Each panel indicates which (line) fluxes were used to create the image. The two left-most panels show the data with original pixelation, while the other panels show the tessellated data (S/N(H$\alpha$)=600). The panels showing emission lines display the image in square root scale, and the continuum image (top-left) is in linear scale. The bottom-right panel shows the 8100-8200\,\AA{} continuum emission in grey, overlaid with orange H$\alpha$ flux contours. Magenta regions indicate the cluster names taken from dG13 for some particular clusters. The black-bordered magenta crosses indicate the positions of the SSCs, from east to west: cluster \#5, the supernebula, and cluster \#11. The black square indicates the position of Complex \#2. In the RGB images, the combination of red and blue gives magenta, the combination of blue and green results in cyan, and the combination of green and red results in yellow. At the position of the composite colours, both components are of equal strength.}
    \label{fig:RGB}
\end{figure*}

Figure \ref{fig:RGB} presents a set of composite images of NGC\,5253 summarising the morphological, ionisation, and extinction structure in the galaxy. This section discusses the set of images. The fluxes used to create the images are from the emission line fits, unless otherwise stated.

The top-left panel in Fig. \ref{fig:RGB} shows the continuum emission from the galaxy. The three colour bands have been chosen in a way to avoid inclusion of (stronger) emission lines. Because of our high spatial resolution, we can resolve individual star clusters. Several clusters in this image have been identified and characterised by dG13, but there are many without identification. The large and bright cluster in the centre of the image is cluster \#5. The group of blue clusters in the southern part of the image is Complex \#2 in MI10, or UV1 in \cite{kobulnicky_hubble_1997} (see the bottom-right panel for labelled cluster names). We note that the difference in colours between the clusters is not only due to different ages, but also because of differences in extinction, making some clusters appear redder than others.

We show the three different hydrogen recombination lines Pa9, H$\alpha$, and H$\beta$ in the second panel of the top row of Figure \ref{fig:RGB}. This image indicates the gas extinction structure in the galaxy, with redder parts of the image corresponding to higher levels of extinction. Already, we can discern a complex extinction structure with clear filaments and clumps. Prominent are the two large bubbles to the north-west and south-east of the SSCs. The western bubble is more filamentary, while the eastern one appears more hollow. There is also a clear difference in extinction between cluster \#5 and cluster \#11. An elaborate discussion on the attenuation in the galaxy can be found in sections \ref{sec:results_attenuation} and \ref{sec:discussion_attenuation}.

The top-right panel displays oxygen emission lines of different levels of ionisation: [\ion{O}{I}]$\lambda$6300, [\ion{O}{II}]$\lambda$7320, and [\ion{O}{III}]$\lambda$5007. From this plot, we can characterise the ionisation structure of the galaxy. There is a region with high ionisation close to the SSCs and Complex \#2 (adhering to the MI10 nomenclature), caused by the ionising photons from the young stars. The inside of the bubbles also consists of gas with a high level of ionisation, while the edges of the bubbles are of intermediate ionisation. The prominent cluster with intermediate ionisation in the north-west is identified as cluster \#105 by dG13 (see the bottom-right panel). Lastly, according to expectation, the gas with lower ionisation is found at larger radii from the SSCs. An exception to this is the region of high ionisation at the north of the image and the region on the south-east. 

The bottom-left panel shows the emission of H$\alpha$, [\ion{N}{II}]$\lambda$6584, and [\ion{O}{III}]$\lambda$5007 as obtained by manually creating narrow-band images at the positions of the line, while removing the contribution from the stellar continuum. Here, we identify an even more complex structure. The eastern bubble also appears to have opened up, showing none of the loops that are visible in the north-west. Strong emission originates from the region surrounding the SSCs, as well as a large number of ionising photons coming from Complex \#2, as visible by the H$\alpha$ and O$^{++}$ emission. Overall, [\ion{O}{III}]$\lambda$5007 and H$\alpha$ show a very similar distribution. [\ion{N}{II}]$\lambda$6584 on the other hand, displays a very different morphology, with bright emission directly to the north of the SSCs. Since this behaviour is not seen for the extinction or ionisation structure, it is already a signal for the presence of nitrogen enrichment. An elaborate discussion on the nitrogen enrichment is given in Sect. \ref{sec:results_abundances} and \ref{sec:discussion_NO}.

The middle panel of the bottom row shows the \ion{He}{I}$\lambda$6678 nebular emission, as well as that of [\ion{S}{II}]$\lambda$6731 and [\ion{S}{III}]$\lambda$9069. The two sulphur lines trace the ionisation structure of the galaxy. S$^{+}$ traces similar structures as O$^0$, while S$^{++}$ shows a morphology that is comparable to a combination of that of O$^{+}$ and O$^{++}$. \ion{He}{I} shows a distribution similar to that of the bulk of the ionised gas traced by H$\alpha$, except for it being stronger on the south-eastern edge of the field of view (FOV).

\subsection{Attenuation} \label{sec:results_attenuation}
The dust distribution in NGC\,5253 is patchy, resulting in a non-uniform attenuation\footnote{With attenuation we refer to the total process of the absorption and scattering of photons both out of and into the line of sight by the dust.}. Correcting for the effects of attenuation is essential to obtain correct measurements of physical properties. Beyond this, it is interesting to map the dust at the parsec scales that are available within the data.

We derived the attenuation of the gas of NGC\,5253 by exploiting the difference between the observed and theoretical hydrogen recombination line ratios \citep{osterbrock_astrophysics_2006}. Within the data, we can access the hydrogen recombination lines H$\alpha$, H$\beta$, and the Paschen series from n=9 to n=16. We limit ourselves to using H$\alpha$, H$\beta$, and Pa9, because the other lines become too low S/N at the quality we aim to perform this analysis. 

\begin{figure*}
    \centering
    \includegraphics[width=\textwidth]{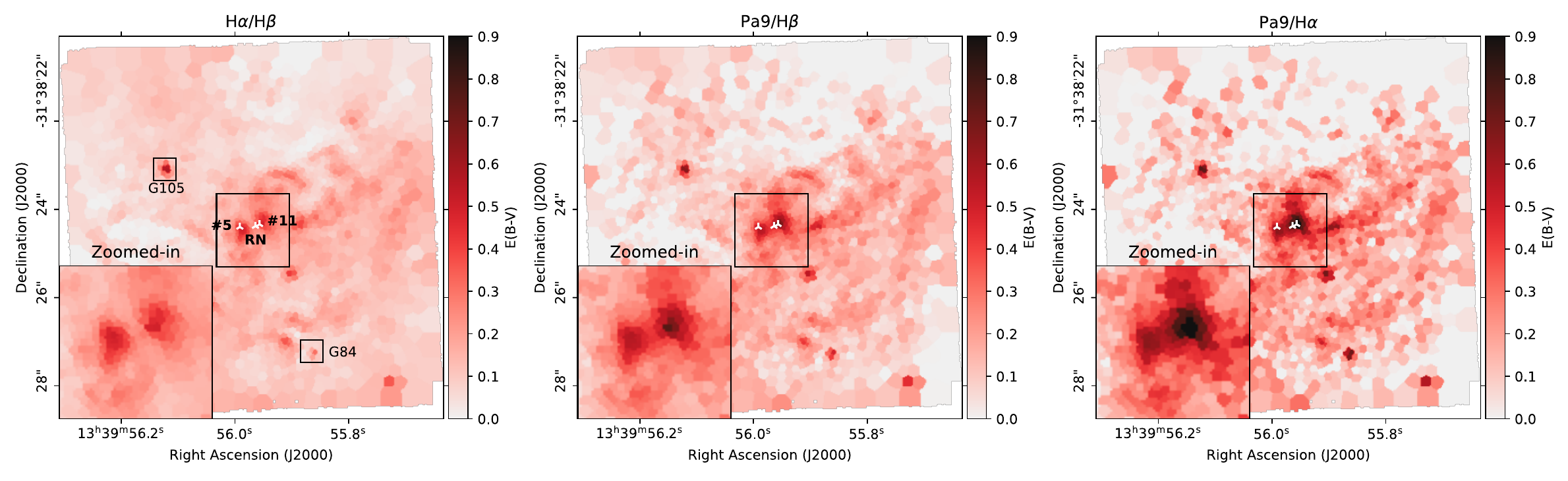} 
    \caption{E(B-V) determined using the line ratios H$\alpha$/H$\beta$ (left), Pa9/H$\beta$ (middle), and Pa9/H$\alpha$ (right). The inset in each panel is a zoom-in on the central region. The positions of cluster \#5, the supernebula, and cluster \#11 are indicated by the white markers (from left to right), positions of dG105 and dG84 are indicated by black squares in the first panel. For these maps, a value of $R_V=3.1$ is used. }
    \label{fig:EBV_allratios}
\end{figure*}

For the line ratios Pa9/H$\alpha$, Pa9/H$\beta$, and H$\alpha$/H$\beta$, we use \texttt{PyNeb}\footnote{We use the default atomic data as of 2025. Table \ref{tab:AtomicData} lists the references for the atomic data used to produce the results throughout this paper.}\citep{luridiana_pyneb_2015} to calculate their theoretical (intrinsic) values. This is done assuming Case B recombination and with an electron temperature of $T_\mathrm{e}=10000$\,K and electron density as determined by FADO ($n_\mathrm{e,mean}=266$\,cm$^{-3}$, $n_\mathrm{e,max}=1590$\,cm$^{-3}$). Because of the dependency of $T_\mathrm{e}$ and $n_\mathrm{e}$ on the assumed attenuation, the attenuation and $T_\mathrm{e}$ and $n_\mathrm{e}$ should in principle be calculated iteratively \citep{ueta_proper_2021}. However, we confirm that the exact choice of $T_\mathrm{e}$ and $n_\mathrm{e}$ has minimal influence on the results. To obtain the reddening (E(B-V)), we compare the obtained theoretical line ratios to their observed values, assuming an attenuation factor of $R_V=3.1$, equal to the Milky Way mean \citep{savage_observed_1979, cardelli_relationship_1989}. For the attenuation law, we use a combination of the average LMC attenuation curve in the UV \citep{fitzpatrick_correcting_1999} and in the optical and the attenuation curve proposed by \cite{fitzpatrick_analysis_1988} for the infrared (\texttt{F99 F88 LMC} in \texttt{PyNeb}). The resulting reddening maps are shown in Figure \ref{fig:EBV_allratios}. We observe significant discrepancies between the three line ratios when examining the three SSCs. Table \ref{tab:EBV_Av} shows the maximum values of reddening for each region. 

These discrepancies indicate that the attenuation law is not a good fit to the data. Thus, we fit for the shape of the attenuation law that is described by the attenuation parameter $R_V$. From observations, it is known that $R_V$ varies between different lines of sight. Toward stars in the Milky Way, values of $2.1\leq R_V\leq5.8$ have been observed \citep{welty_ultraviolet_1992, fitzpatrick_correcting_1999, 2025Sci...387.1209Z}, while in gravitationally lensed galaxies, values as extreme as $1.5\leq R_V\leq7.2$ have been found \citep{falco_dust_1999}.
In this work, we allow for values of $1\leq R_V \leq 13$ and keep the attenuation law fixed to \texttt{F99 F88 LMC}. We fit for $R_V$ -- and consequently the shape of the attenuation law -- for regions where the Pa9 S/N (flux/standard deviation of the continuum) exceeds 100. For each pixel with a high S/N, we compute the E(B-V) values for each of the three line ratios -- Pa9/H$\alpha$, Pa9/H$\beta$, and H$\alpha$/H$\beta$ -- considering different values of $R_V$. Subsequently, we conduct a tile-wise comparison of the E(B-V) values obtained from the three different line ratios. To quantify consistency, we compare E(B-V) values for all possible combinations of two ratios, determining the $R_V$ value that minimises the absolute differences in E(B-V) (normalised by the average value at that tile) for each combination of line ratios. This approach is comparable to that of \cite{rogers_spectral_2024}. 
We estimate uncertainties through 200 Monte Carlo simulations, varying the initial line fluxes according to their fitting errors. The error is then the standard deviation of the resulting distribution. Typically, for the low-S/N area -- where we keep $R_V$ fixed to 3.1 -- the mean error on E(B-V) is found to be 0.08. While for the high-S/N region, we find a mean error on E(B-V) of 0.03 and on $R_V$ of 1.25. 

\begin{figure*}
    \centering
    \includegraphics[width=\textwidth]{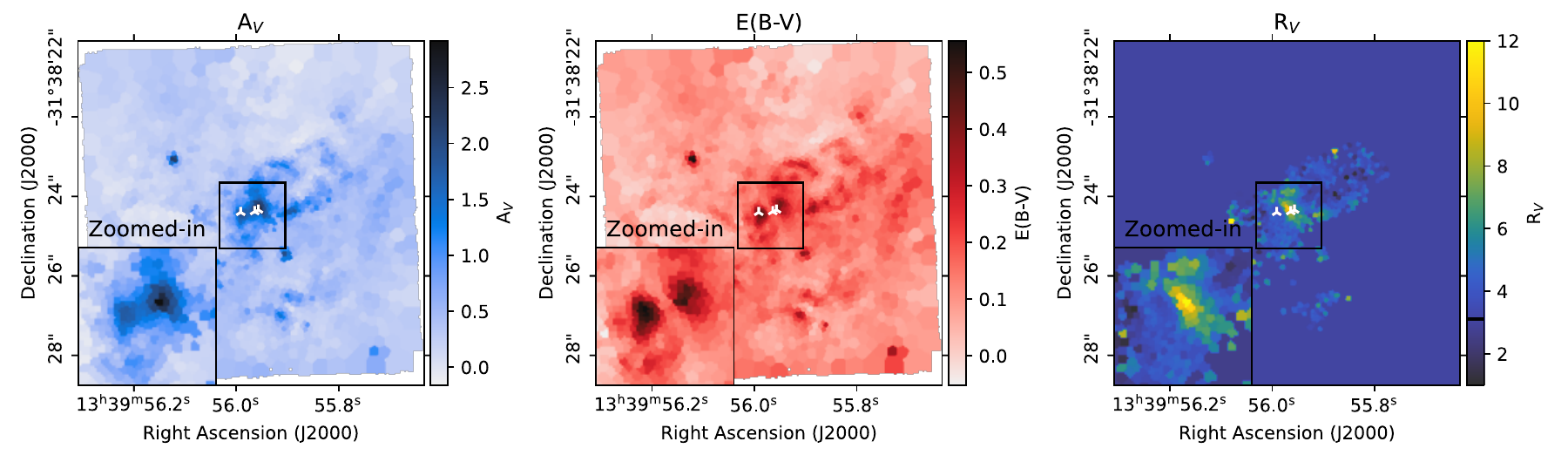} 
    \caption{Visual extinction (left) and reddening (middle) maps for NGC 5253 and the value of $R_V$ used to obtain these (right). The markers in white indicate the position of, from left to right, cluster \#5, the supernebula, and cluster \#11. The inset is the zoom-in on the central part enclosed by the black lines.}
    \label{fig:AV_EBV_Rv}
\end{figure*}

Figure \ref{fig:AV_EBV_Rv} shows the resulting maps of the visual extinction $A_V=E(B-V)\times R_V$, E(B-V), and $R_V$. The filamentary structure of the H$\alpha$ emission is also visible in the map of E(B-V). The reddening and visual extinction values for the central area can be found in Table \ref{tab:EBV_Av}. Due to the differences in $R_V$, a larger E(B-V) now no longer necessarily corresponds to a larger $A_V$ when comparing between different clusters. We find lower values of reddening and visual extinction for the supernebula than for clusters \#5 and \#11, which indicates that we are only tracing a small fraction of the cloud in which the cluster is embedded. Interestingly, we find significantly higher values of the extinction parameter for the supernebula of $R_V=9.67\pm1.45$. These results and their physical implications are further discussed in Sect. \ref{sec:discussion_attenuation}.

We identify several additional areas of significant reddening across the galaxy. The region showing high values of reddening (E(B-V)=0.53) in the north-east of the galaxy (13h39m56.1s, -31$^\circ$38'23.0") coincides with cluster \#105 characterised by dG13 (hereafter referred to as G105) to have an age of log(t/yr)=6.3. The young age of this cluster is compatible with the high level of attenuation at that position. Another region of high reddening is for the cluster at position 13h39m55.86, -31$^\circ$38'27.24" (south-west), which corresponds to cluster \#84 from dG13 (hereafter dG84). The age for this cluster is not available in the literature. The spectrum of the source shows a very broad component in H$\alpha$, suggestive of it being a Wolf-Rayet star. 

\begin{table*} 
    \centering
    \begin{tabular}{lccccc} 
        \thickhline
      & H$\alpha$/H$\beta$ & Pa9/H$\beta$ & Pa9/H$\alpha$ & Vary $R_V$ & $R_V$ \\
     \thickhline
     E(B-V) cluster \#5 & 0.50 $\pm$ 0.03 & 0.56 $\pm$ 0.02 & 0.64 $\pm$ 0.04 & 0.48 $\pm$ 0.05 & 3.89 $\pm$ 0.51\\
     E(B-V) supernebula & 0.29 $\pm$ 0.03 & 0.53 $\pm$ 0.02 & 0.84 $\pm$ 0.04 & 0.22 $\pm$ 0.03 & 9.67 $\pm$ 1.45\\
     E(B-V) cluster \#11 & 0.49 $\pm$ 0.03 & 0.76 $\pm$ 0.02 & 1.10 $\pm$ 0.04 & 0.42 $\pm$ 0.05 & 6.89 $\pm$ 0.66\\


    \thickhline
     $A_V$ cluster \#5 & 1.56 $\pm$ 0.12 & 1.75 $\pm$ 0.06 & 1.98 $\pm$ 0.15 & 1.88 $\pm$ 0.10 & 3.89 $\pm$ 0.51\\
     $A_V$ supernebula & 0.88 $\pm$ 0.10 & 1.65 $\pm$ 0.07 & 2.61 $\pm$ 0.15 & 2.17 $\pm$ 0.10 & 9.67 $\pm$ 1.45\\
     $A_V$ cluster \#11 & 1.51 $\pm$ 0.09 & 2.35 $\pm$ 0.05 & 3.40 $\pm$ 0.12 & 2.91 $\pm$ 0.08 & 6.89 $\pm$ 0.66\\

     \thickhline
    \end{tabular}
    \caption{Value of E(B-V) and $A_V$ for cluster \#5, the supernebula, and cluster \#11. The columns show the results using different line ratios, the two rightmost columns show the best fit result for when we fit for $R_V$ and the corresponding best-fit $R_V$, all other results are for $R_V=3.1$. The errors are based on 200 Monte Carlo simulations and correspond to the standard deviation of the resulting distribution.}
    \label{tab:EBV_Av}
\end{table*}

\subsection{Mapping electron temperature and density} \label{sec:results_Te_ne}
In this section, we discuss the electron temperatures ($T_\mathrm{e}$) and densities ($n_\mathrm{e}$) in NGC\,5253. The spatially resolved determination of these properties is important for the precise determination of the abundances, done in Section \ref{sec:results_abundances}. 

$T_\mathrm{e}$ and $n_\mathrm{e}$ are mapped using the line ratios of collisionally excited lines. The line ratios used are dependent on either $T_\mathrm{e}$ or $n_\mathrm{e}$, while being (nearly) independent of the other. This is achieved by choosing ions that, in the case of determining $T_\mathrm{e}$, have at least two (upper) energy levels with considerably different excitation energies. Or, for $n_\mathrm{e}$, comparing two lines that have approximately the same excitation energies so that the excitation rates depend only on the ratio of the collision strengths \citep{osterbrock_astrophysics_2006}.

 Specifically, we use [\ion{N}{II}]$\lambda$5755/(6548+6584) (hereafter $T_\mathrm{e}$([\ion{N}{II}])) and [\ion{S}{II}]$\lambda$6731/6716 (hereafter $n_\mathrm{e}$([\ion{S}{II}])) for $T_\mathrm{e}$ and $n_\mathrm{e}$ (resp.) in a low ionisation plasma. For the high ionisation plasma, we characterise $T_\mathrm{e}$ via [\ion{S}{III}]$\lambda$6312/9069 (hereafter $T_\mathrm{e}$([\ion{S}{III}]))\footnote{Because of the lower ionisation energy of 34.78\,eV for S$^{++}$, than for than for O$^{++}$, which is used by many authors, of 54.9\,eV, the plasma we characterise here will be referred to by many authors as a medium ionisation plasma.}. The electron density can be defined by [\ion{Cl}{III}]$\lambda$5538/5518 (hereafter $n_\mathrm{e}$([\ion{Cl}{III}])). However, the S/N of these lines is too low to map the density, therefore, we fix $n_\mathrm{e}$([\ion{Cl}{III}])=4000\,cm$^{-3}$, the median for the tiles with an error $<$50\%.

To determine the electron temperatures and densities, we make use of \texttt{PyNeb}'s \texttt{getTemDen} method in its \texttt{Atom} class. We assume a fixed, but reasonable, temperature or density while calculating the other quantity from the relevant line ratio. We treat the high and low ionisation plasma separately. The best fit $T_\mathrm{e}$ and $n_\mathrm{e}$ are the median value from 1000 Monte Carlo simulations, and the errors correspond to the 16th and 84th percentiles of the same distribution

\begin{figure*}
    \centering
    \includegraphics[width=\linewidth]{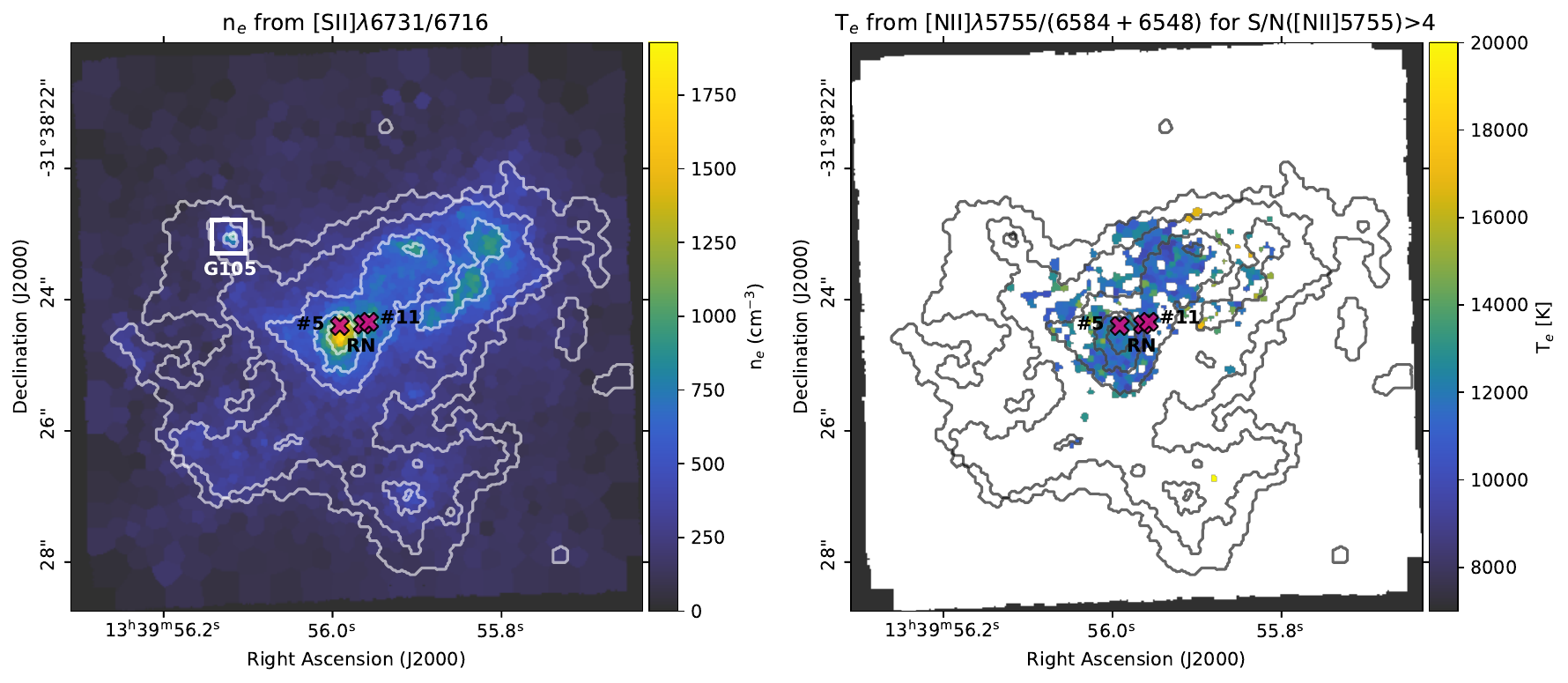}
    \caption{Electron density (left) and temperature (right) for the low ionisation plasma as determined by the line ratios stated in the respective titles. The magenta markers indicate the position of, from left to right, cluster \#5, the supernebula, and cluster \#11. The white square indicates the position of cluster G105. The white/black contours are of the H$\alpha$ flux. For future calculations, tiles with S/N([\ion{N}{II}]$\lambda$5755)$<$4 will use the median value of T$_e$([\ion{N}{II}]) = 11842\,K derived from the high-S/N tiles. }
    \label{fig:Tene_low_Vorbin}
\end{figure*}

Figure \ref{fig:Tene_low_Vorbin} shows the determined electron temperature for the low-ionisation plasma. We obtain the values for $n_\mathrm{e}$([\ion{S}{II}]) by fixing $T_\mathrm{e}$ to 12000\,K, which is a typical value for $T_\mathrm{e}$([\ion{N}{II}]) for our data. This results in the $n_\mathrm{e}$([\ion{S}{II}]) measurements shown on the left-hand side of Figure \ref{fig:Tene_low_Vorbin}. We find that the variations in $n_\mathrm{e}$([\ion{S}{II}]) follow the structure of the ionised gas, traced by H$\alpha$ emission, very well. The maximum value of $n_\mathrm{e}=1930\pm40$\,cm$^{-3}$ is found slightly south of the position of cluster\,\#5. A second maximum value is found 0\farcs1 to the south of this first maximum, with a value of $n_\mathrm{e}=1796\pm50$\,cm$^{-3}$. The distance between the two peaks in density is at the limit of what we can resolve, with a median FWHM of the PSF measured at 0\farcs15 (see Section \ref{sec:resolution}). The location of the second peak in $n_\mathrm{e}$ coincides with an area of diffuse stellar emission, but no resolved cluster is detected (neither by us nor in the literature). We also find an enhancement in $n_\mathrm{e}$, with a value of $n_\mathrm{e}=910\pm22$\,cm$^{-3}$ at the location of cluster G105. This clearly shows that we find variations of $n_\mathrm{e}$ at the scale of individual clusters. 

The electron temperature, $T_\mathrm{e}$([\ion{N}{II}]), is determined using the best-fit $n_\mathrm{e}$([\ion{S}{II}]) values. The measurements are noisy due to large uncertainties in the flux measurements of the faint auroral line [\ion{N}{II}]$\lambda$5755. Fitting of this line is further complicated because it is located near the spectral region affected by AO-laser masking and the onset of the "red bump" from Wolf-Rayet stars. Therefore, we restrict the analysis to tiles with S/N([\ion{N}{II}]$\lambda$5755)$>$4, which reduces the dataset to 393 tiles out of 2149. These tiles are primarily concentrated around and north-west of the SSCs. We consider tiles with S/N([\ion{N}{II}]$\lambda$5755)$<$4 to have the median value of $T_\mathrm{e}$([\ion{N}{II}]) derived from the high-S/N tiles of $T_\mathrm{e}=11842 \pm 1667$\,K (median $\pm$ standard deviation over the tiles). Adopting this approach assumes that $T_\mathrm{e}$ is uniformly distributed across the FOV. Although this may not strictly be the case, the $T_\mathrm{e}$ distribution appears relatively uniform in regions with high S/N measurements.

\begin{figure}
    \centering
    \includegraphics[width=\linewidth]{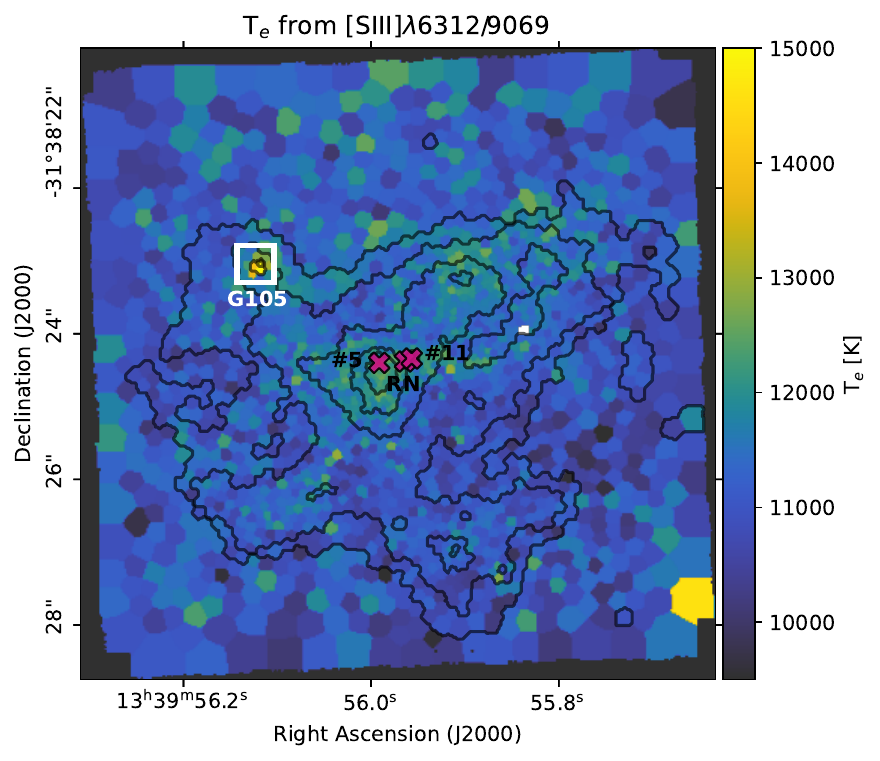} 
    \caption{Electron temperature from [\ion{S}{III}]$\lambda$6312/9069, with $n_\mathrm{e}$ fixed to $n_\mathrm{e}$=3986 cm$^{-3}$. The magenta markers indicate the position of, from left to right, cluster \#5, the supernebula, and cluster \#11. The white square indicates the position of cluster G105. The black contours trace the H$\alpha$ flux.}
    \label{fig:Temedian}
\end{figure}

To calculate the electron temperature for the high ionisation plasma, we fix the electron density at $n_\mathrm{e}=4000 \text{ cm}^{-3}$, the median value for tiles with upper errors smaller than 50\% of the best-fit density. Fortunately, the exact choice of $n_\mathrm{e}$ has a negligible impact on the temperature measurements. The best-fit $T_\mathrm{e}$ value is then determined as the median of 1000 simulations. The resulting temperature map is shown in Figure \ref{fig:Temedian}. Generally speaking, the map is relatively flat, with a median $T_\mathrm{e}$ of 11260$\pm$575 K, but it also reveals some notable structures. One such structure is the enhancements in $T_\mathrm{e}$ around the SSCs and the northeastern bubble, though these features are weak. The most prominent structure is the sharp increase in $T_\mathrm{e}$ to $15400^{+110}_{-130}$ K at the position of G105. This elevated temperature may be consistent with the cluster’s young age. However, considering the uncertainties in both $T_\mathrm{e}$ and extinction, it could also be compatible with a flat temperature distribution. To determine if the increase is real, deeper observations are necessary.

\subsection{Chemical abundances} \label{sec:results_abundances}
Now that we have performed a correction for extinction and the electron temperatures and densities have been determined, it is possible to determine the ionic and atomic abundances. We quantify the abundances of helium, oxygen, and nitrogen. 

\subsubsection{Helium abundance}\label{sec:HeliumAbundance}
We use \texttt{PyNeb}'s \texttt{RecAtom} class to determine the abundance of singly ionised helium, y$^+$. We adopt $T_\mathrm{e}$([\ion{S}{III}]) and $n_\mathrm{e}$([\ion{S}{II}]) as the electron temperature and density, respectively, for determining the theoretical emissivity. The exact choice of $T_\mathrm{e}$ and $n_\mathrm{e}$ has a negligible influence on the final abundances. Because the emission coefficients for recombination lines are not very sensitive to temperature, scaling approximately as $\epsilon_l\propto 1/T$. 

At the position where dG13 detect cluster G106, we detect a very prominent broad component in \ion{He}{I}$\lambda$6678, possibly caused by a WR-star. In this region, we fit a double Gaussian to the emission lines and use only the narrow component to determine the abundance. 

\begin{figure}
    \centering
    \includegraphics[width=0.95\linewidth]{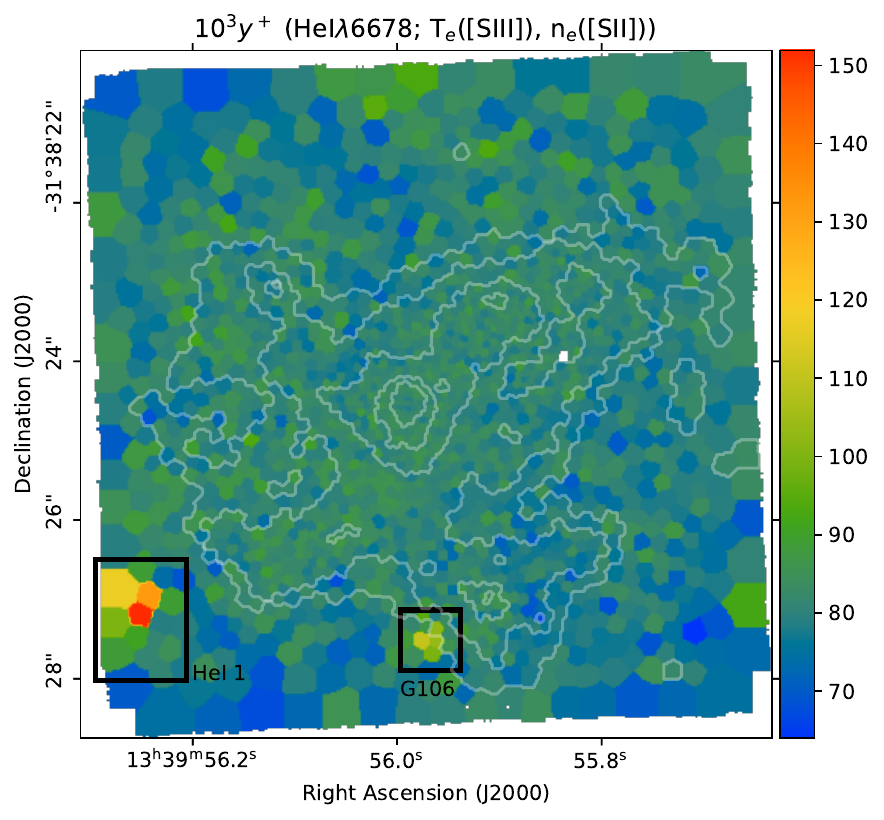}
    \caption{Map of $10^3 y^+$ derived for the main component of \ion{He}{I}$\lambda$6678. $T_\mathrm{e}$([\ion{S}{III}]) an $n_\mathrm{e}$([\ion{S}{II}]) are used. The white contours trace the emission in H$\alpha$. The magenta squares indicated regions of specific interest (see text).}
    \label{fig:abundHeI6678}
\end{figure}

The resulting abundances for \ion{He}{I}$\lambda$6678 can be found in Figure \ref{fig:abundHeI6678}. The map is flat, with a mean abundance ($\pm$ standard deviation) of $10^3y^+=80.9 \pm 4.3$. This is consistent with the results of \cite{monreal-ibero_he_2013}, who obtain a mean of $10^3y^+=76.8\pm1.8$ for \ion{He}{I}$\lambda$6678. 

We identify two regions of particular interest, marked by the black boxes in Figure \ref{fig:abundHeI6678}. 
The first region of interest is G106. Where, even in the narrow component of the fit, we detect an enhancement in $y^+$. For the broad component, we detect a value of $10^3y^+$ of maximum 100. This is considering a normalisation with the broad component of H$\beta$ and with the same $T_\mathrm{e}$ and $n_\mathrm{e}$ as for the narrow component. \cite{monreal-ibero_he_2013} did not detect such an increase in $y^+$ abundance for this region, likely due to their lower spatial resolution. 

The second region of enhanced \ion{He}{I} emission, labelled HeI\,1, also shows elevated $y^+$. The location of the enhancement in $y^+$ aligns with a cluster detected in the MUSE data (see Fig. \ref{fig:RGB}), but not catalogued by dG13. The matching positions of the cluster and the enhancement is suggestive of the cluster being the cause of the enhancement. However, contrary to G106, we do not detect a broad component for H$\alpha$ or \ion{He}{I} in its spectrum. A more detailed characterisation of this cluster is necessary to determine the exact relation between the cluster and the \ion{He}{I} enhancement.

\subsubsection{Oxygen abundance}
To calculate the abundance of oxygen, observed through collisionally excited lines, we make use of the \texttt{PyNeb} method \texttt{getIonAbundance} for their \texttt{Atom} class. We make use of the default atomic data (see Appendix \ref{sec:appendix_atomicdata}). Importantly, for collisionally excited lines, the emission coefficients scale exponentially with temperature. Thus, the determined abundance is extremely sensitive to the chosen electron temperature.  

We determine the ionic abundances for the ions O$^0$, O$^+$, and O$^{++}$ through the transitions [\ion{O}{I}]$\lambda$6300, [\ion{O}{I}]$\lambda$6364, [\ion{O}{II}]$\lambda$7320 (a doublet), [\ion{O}{II}]$\lambda$7331 (a doublet), [\ion{O}{III}]$\lambda$4959, and [\ion{O}{III}]$\lambda$5007. All line fluxes are corrected for extinction and normalised to H$\beta$. For O$^0$ and O$^{+}$, we use the temperatures and densities belonging to the low-ionisation plasma. Because $T_\mathrm{e}$([\ion{N}{II}]) is very noisy, even for the high-S/N  pixels, we use the median value of $T_\mathrm{e}=11842$\,K for the whole FOV. Doing so smooths our final results but does not lead to any large differences in the determined abundance structure and values. For O$^{++}$ we use $T_\mathrm{e}$([\ion{O}{III}]) as obtained from the $T_\mathrm{e}([\ion{O}{III}])-T_\mathrm{e}([\ion{S}{III}])$ relation provided by \cite{brazzini_metallicity_2024}. 

The resulting ionic abundances show a morphology that is expected based on the ionisation structure of the galaxy (see Fig. \ref{fig:RGB}). This means that in the regions where the level of ionisation is low, we find a relatively high abundance of O$^0$, while in regions of high ionisation, there is a high abundance of O$^{++}$ and very little neutral O$^0$ and O$^+$. The maps of the abundances for the individual transitions can be found in Fig. \ref{fig:IonicOxygen}. 

\begin{figure}
    \centering
    \includegraphics[width=\linewidth]{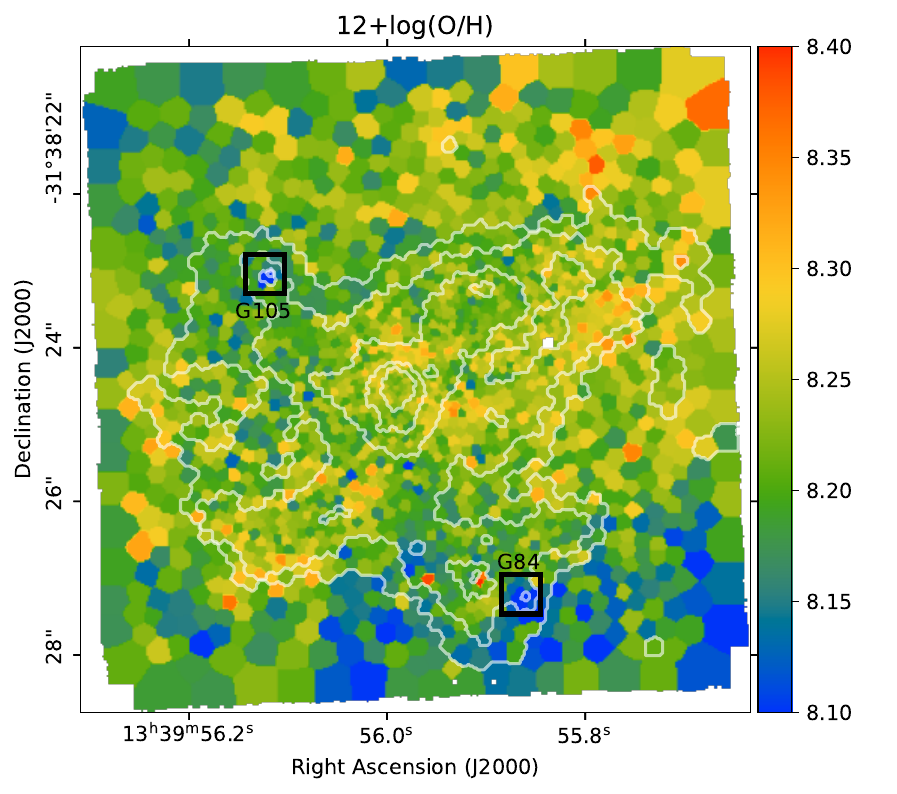}
    \caption{Total oxygen abundance 12+log(O/H). The white contours indicate the H$\alpha$ flux. The black squares mark the position of clusters G105 and G84.}
    \label{fig:logOH}
\end{figure}

To obtain the total oxygen abundance we use 12+log(O/H)\,=\,12+log((O$^{+}$ + O$^{++}$)/H$^{+}$) \citep{stasinska_abundance_2002}. We weigh the abundances of each ion by the emissivity of each transition. The resulting map of oxygen abundance is presented in Fig. \ref{fig:logOH}. The distribution is relatively flat, with a mean $\pm$ standard deviation of 12+log(O/H)=8.22$\pm$0.05. These measurements agree well with the values from literature \citep{monreal-ibero_ionized_2012, lopezsanchez_localized_2007}.

We find a lower oxygen abundance at the positions of G105 and G84 of 12+log(O/H)=8.0 in both cases. This is inconsistent with a homogeneous oxygen abundance distribution at this location, even when considering the errors in extinction and temperature. In fact, the electron temperature needs to be as low as $T_e$([\ion{S}{III}])$\approx$8400\,K for G105 to have an oxygen abundance equal to the median value. This is not only much lower than the median electron temperature, but it is especially inconsistent with the observed increase in $T_e$ at that location. In Sect. \ref{sec:results_Te_ne} we argued that the observed increase in $T_e$([\ion{S}{III}]) could potentially be explained by the high extinction at the position of G105. Whilst an overestimation of extinction indeed results in an underestimation of total oxygen abundance, the total amount of extinction will need to be near-zero at the position of G105 and G84 to be consistent with a flat oxygen abundance distribution. Therefore, we argue that we are seeing the actual existence of small pockets of less enriched gas at the location of these clusters. 

\subsubsection{Nitrogen abundance}\label{sec:results_N}
We determine the ionic abundances of nitrogen from [\ion{N}{II}]$\lambda$6584, with an approach identical to that for oxygen. From there, we calculate the total nitrogen abundance relative to that of oxygen via the simple expression $N/O = N^+/O^+$ \citep{stasinska_abundance_2002}, where we assume the ICF=1 because of the similar ionisation energies of N$^+$ and O$^+$. 

\begin{figure}
    \centering
    \includegraphics[width=\linewidth]{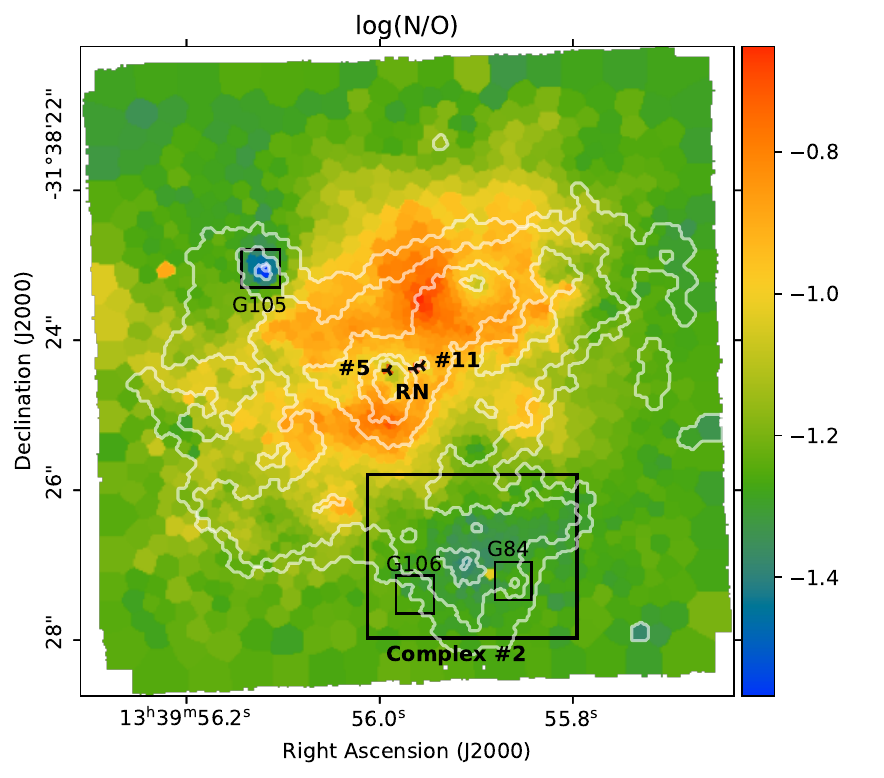}
    \caption{log(N/O) determined with $T_\mathrm{e}=11842$\,K. White contours follow the H$\alpha$ flux. The black markers indicate the positions of cluster \#5, the supernebula, and cluster \#11, while the black squares show the positions of clusters G84, G105, G106 and Complex \#2.}
    \label{fig:logNO}
\end{figure}

Figure \ref{fig:logNO} shows the resulting nitrogen abundance. For the first time, we are able to map the nitrogen enhancement at a spatial resolution of 2.3 pc. We clearly observe the factor 2-3 nitrogen enhancement as compared to the median N/O in the FOV, which is in agreement with the literature. The area containing N-enriched material presents a different morphology than that of the bulk of the ionised gas as traced by H$\alpha$. Instead, the largest level of enhancement is observed north of the SSCs. Interestingly, we observe a deficit of nitrogen with respect to oxygen at the position of G105. This could be a result of our assumption of a flat $T_\mathrm{e}$, because a local increase in temperature of $T_\mathrm{e}\sim4000$\,K will erase the inhomogeneity. Such an increase in temperature is not unreasonable because such a high $T_\mathrm{e}$ is detected for the higher ionisation plasma (see Fig. \ref{fig:Temedian}), but impossible to measure at the depth of our data because of the faintness of the [\ion{N}{II}]$\lambda5755$ line. We note that the low oxygen abundance we observe at this same position cannot explain the low value of N/O. Instead, if the oxygen abundance were the same as the median O/H, this would only result in a much lower N/O.

Almost no enhancement is found for Complex \#2. In the area directly surrounding the SSCs we also find relatively little enhancement, whereas we do observe a clear enhancement in N/O in the eastern bubble, extending outward to the edge of our FOV. Section \ref{sec:discussion_NO} will discuss in great detail the possible origin of the extra nitrogen.

\subsection{Wolf-Rayet stars}
One of the most popular theories to explain the extra nitrogen in NGC\,5253 is the presence of Wolf-Rayet (WR) stars \citep{walsh_optical_1989, kobulnicky_hubble_1997,lopezsanchez_localized_2007}. In this section, we present a complete census of the position and quantity of WR stars within our FOV at an unprecedented level of detail. This is done by combining the WFM data with results from MI10. Section \ref{sec:discussion_NO} will discuss in detail the relation between the presence of WR stars and the nitrogen enrichment.

WR stars descend from O-type stars. They exhibit very strong and broad emission lines caused by their powerful stellar winds. The phase is short-lived, lasting $\sim$1\,Myr \citep{1994A&A...287..803M}. Broadly speaking, there are two main types of WR stars: the WN subtype, showing products of the CNO cycle and its evolution; and the WC subtype, which shows strong He, C, and O, and is a product of the triple-$\alpha$ process \citep{crowther_physical_2007}. Observationally, WR stars can be identified in distant -- i.e. where stars are not spatially resolved -- galaxies by the so-called blue and red WR bumps. The blue WR bump is the most prominent feature, between 4650-4670\AA, caused by an unresolved blend of nitrogen, carbon, and helium lines originating from WN-type WR stars. The red WR bump originates from a blend of broad carbon emission lines in the atmospheres of WC-type WR stars. Unfortunately, the wavelength coverage of the NFM makes it impossible to detect either of these features. However, the WFM covers the red bump. 

\begin{figure}
    \centering
    \includegraphics[width=\linewidth]{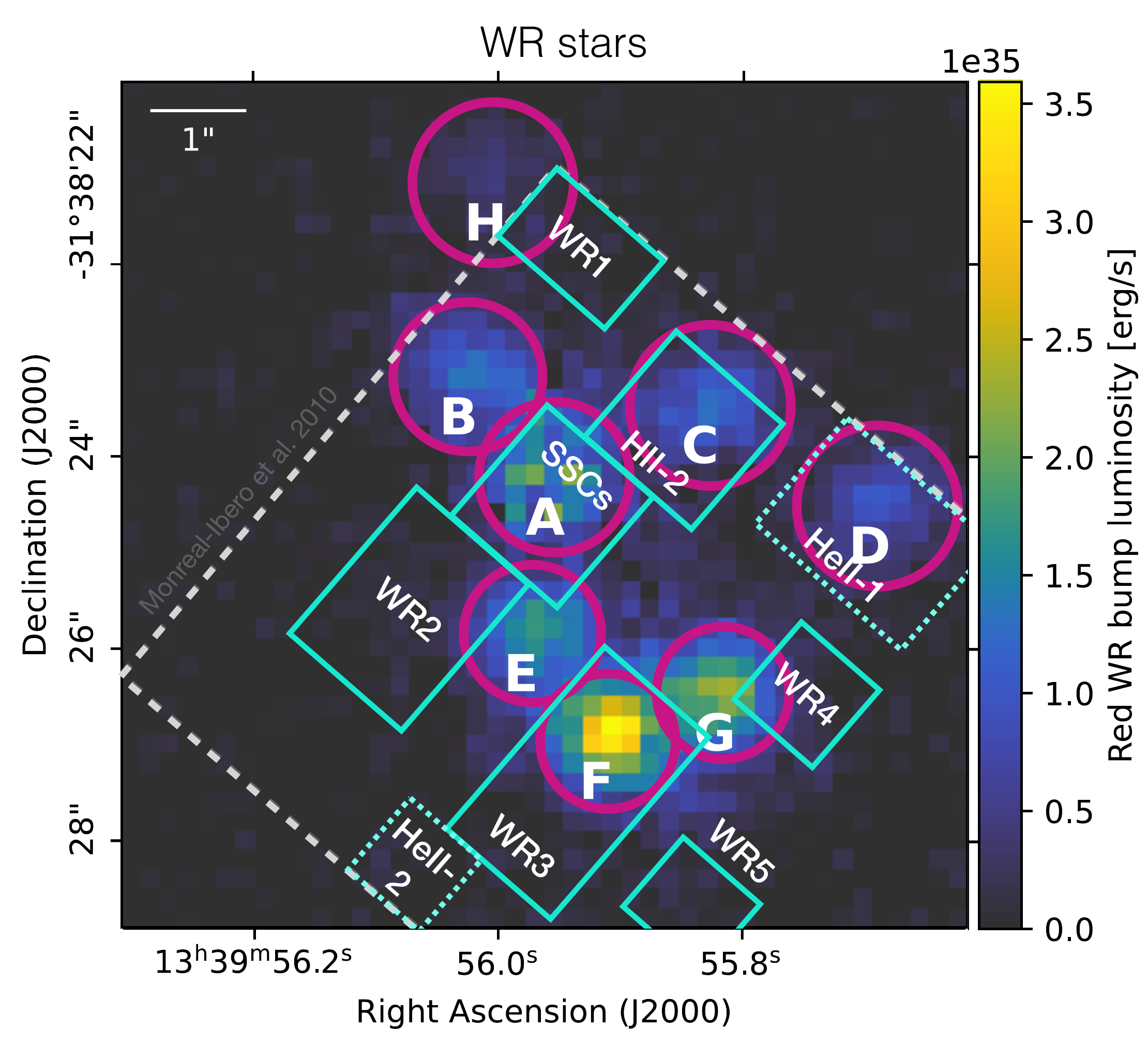}
    \caption{Luminosity map of the red WR bump. The magenta circles indicate regions where the red bump was detected, and the letters inside identify the name of the region. The solid cyan rectangles are the regions where MI10 detect the blue WR bump. The dashed cyan rectangles are where they detect \ion{He}{II} emission but not the blue bump. The MI10 FOV is indicated with the grey dashed region. Estimates of the number of WR enclosed in each region can be found in Table \ref{tab:WRcounts}.}
    \label{fig:numWRstars}
\end{figure}

To map the presence of the WR stars, we measure the (reddening-corrected) flux in the window between 5765-5870\AA, where the red bump is located, and subtract the average of the continuum level in two windows on either side of the bump (with a width of $\sim$200\AA). The integrated map of the red WR bump can be found in Fig. \ref{fig:numWRstars}. Several regions with WC stars can be identified. The regions have a diameter of $\sim0\farcs8$, which is consistent with them being marginally resolved at best, given that the WFM seeing FWHM $\sim1$\arcsec. 

We estimate the total number of WC-type stars from the predicted luminosity of a single WR star \citep{crowther_reduced_2006, lopez-sanchez_massive_2010}, using equation (8) from \cite{lopez-sanchez_massive_2010}:
\begin{equation}
    L_{\mathrm{WCE}}(\ion{C}{IV}\,\lambda5808) = (-8.198+1.235x)\times10^{36}\,\text{erg s}^{-1}
\end{equation}
where $x=$12+log(O/H). Assuming our mean metallicity of 12+log(O/H)=8.2, we find that the \ion{C}{IV}$\lambda$5808 luminosity of a single star is $1.9 \times 10^{36}$ erg s$^{-1}$. Table \ref{tab:WRcounts} lists the number of WR stars in every region with a clear red bump feature. Most regions are consistent with encompassing 1-4 WC stars, consistent with the findings of \cite{westmoquette_piecing_2013}. 

Within the FLAMES IFU data, MI10 were able to detect the blue WR bump, as well as areas with nebular \ion{He}{II} but no blue WR bump. The regions identified by MI10 can be found as the cyan rectangles in Fig. \ref{fig:numWRstars}. Most of their regions overlap with the locations where we detect the red bump. The discrepancy in exact positions can possibly be attributed to the less accurate astrometry in the FLAMES data.

We also make a rough estimation of the number of WN stars in every region. To obtain this, we use the estimated log(WR/(WR+O)) from MI10 in combination with the relation from \cite{schaerer_new_1998} and the measured H$\beta$ flux from the WFM data, to estimate the blue bump flux in the regions identified by MI10. We then convert the flux to a luminosity and compare it to the theoretical luminosity of the blue bump for a single WN star at a metallicity of 12+log(O/H)=8.2: $L_{WNL}=1.2\times10^{36}$\,erg\,s$^{-1}$ \citep{lopez-sanchez_massive_2010}. We find between 1-7 WN stars per region. This leads to a total count of $\sim$24 WN stars. The number of WN stars found in each region is listed in Table \ref{tab:WRcounts}. 

In total, we estimate that there are approximately 40 WR stars (both WC and WN types) in the central region of NGC\,5253. The ratio of WC/WN is 0.7$_{-0.3}^{+0.3}$, this value is higher than model predictions by \cite{meynet_stellar_2005} and \cite{eldridge_effect_2008} of 0.1-0.2. However, our estimate likely carries larger uncertainties than those reported here, as these do not include the uncertainty in the luminosity of individual WR stars. Furthermore, while such a high ratio is peculiar, Examples of comparable ratios in similar galaxies in the literature do exist. The BCD galaxy IC10 has a WC/WN ratio of 1.0 \citep{tehrani_revealing_2017}, which the authors attribute to the high star formation intensity (SFR$_{H\alpha}$=0.045\,M$_\odot$yr$^{-1}$) observed in the galaxy. Interestingly, this galaxy also shows signs of nitrogen enhancement \citep{lopez-sanchez_ionized_2011}, though not as strong as NGC\,5253.

\begin{table}
    \centering
    \begin{tabular}{c c c c c c}
    \thickhline
    Region & Number WC&&&Region & Number WN\\
    \thickhline
        A & 2-3 & &  & SSCs & 3-4\\
        B & 1-2 & &  & - & -\\
        C & 1-2 & &  & HII-2 & 2-3\\
        D & 1-2 & &  & HeII-2 & -\\
        E & 2-3 & &  & WR2 & 7\\
        F & 3-4 & &  & WR3 & 6\\
        G & 2-3 &&  & WR4 & 1-2\\
        H & 0-1 &&  & WR1 & 2\\
        - & - &&  & WR5 & 0-1\\
        \thickhline
        Total & 12-20 &  &  & Total & 21-25\\
    \thickhline
    \end{tabular}
    \caption{Estimate of the number or WR stars in each region indicated in Fig. \ref{fig:numWRstars}. The left-most column indicates the regions with a detection of the red bump, the third column gives the closest region with a detected blue bump.}
    \label{tab:WRcounts}
\end{table}

\section{Discussion} \label{sec:discussion}
\subsection{Attenuation} \label{sec:discussion_attenuation}
\subsubsection{Comparison to literature}
We compare our results from Section \ref{sec:results_attenuation} with those of \cite{calzetti_brightest_2015} (C15), who analysed dust attenuation in clusters \#5 and \#11 using a foreground dust screen model and H$\alpha$/P$\beta$ and H$\alpha$/P$\alpha$ line ratios. They find E(B–V)$_{\#5,C15}$ = 0.48 $\pm$ 0.19 and $A_{V,C15} = 1.5$ mag for cluster \#5, well in line with our results (see Table \ref{tab:EBV_Av}). For cluster \#11 they obtain E(B–V)$_{\#11,C15} = 1.51 \pm 0.26$ and $A_{V,C15} = 4.7$, which is higher than our results of E(B–V)$_{\#11} = 0.42 \pm 0.05$ and $A_{V} = 2.91 \pm 0.08$. The higher attenuation values for cluster \#11 in C15 are expected, given that P$\alpha$ and P$\beta$ probe deeper into dusty regions than Pa9. The agreement for cluster \#5 suggests the dust layer is thin enough for Pa9 to remain optically thin.

C15 argue that a foreground screen cannot explain the observed properties in cluster \#11 and instead propose a model in which dust, stars, and gas are uniformly mixed. This yields a significantly higher reddening of E(B–V) = 15.7$^{+6.7}_{-4.2}$ from SED fitting, which is an even larger discrepancy from our results. Despite this disagreement, both studies indicate distinct dust properties between the two clusters.

\subsubsection{Cluster emergence}
We propose that the observed differences among the SSCs reflect an evolutionary sequence in their emergence from their natal gas clouds. The supernebula appears to be still fully embedded in its natal cloud, reinforced by the observation of CO(3-2) \citep{turner_alma_2017} at the position of this cluster. This would correspond to it being a Type \#2 (clump-\ion{H}{II} region complexes) in the classification of \cite{sun_hidden_2024}. Clusters \#5 and \#11 have no detection of CO(3-2) and are visible in our data. This suggests they are older than the supernebula. Within the same framework, this would place them at Type \#3 (clump-\ion{H}{II} region-cluster complexes), or possibly Type \#4 (exposed \ion{H}{II} region-cluster complexes) in the case of cluster \#5, that is likely in a further stage of emergence as indicated by its lower attenuation. This is also compatible with the ages of the clusters. The supernebula has an age of $<1$\,Myr \citep{smith_three_2020}, while clusters \#5 and \#11 have ages of $1\pm1$\,Myr. Given that the typical duration of the stages (types) is estimated around 1-2\,Myr \citep{sun_hidden_2024}, and the total time for a cluster to emerge from its natal birth cloud $\sim$5-10\,Myr \citep[e.g.][]{knutas_feast_2025}, this is consistent with none of the clusters being fully emerged but possibly in different stages of emergence. 

\subsubsection{Physical properties of the dust}
Larger values of the extinction parameter ($R_V>3.1$) have been observationally associated with regions of higher extinction per cloud \citep[e.g.][]{salim_dust_2018,salim_dust_2020,green_dust_2024,zhang_dust_2024,2025Sci...387.1209Z}. Thus, higher values of $R_V$ are thought to trace larger dust grains, as denser environments facilitate grain growth through accretion and coagulation \citep{cardelli_relationship_1989,weingartner_electronion_2001,draine_interstellar_2003}. This is further confirmed by modelling \citep{dartois_spectroscopic_2024}.
Our results align with this understanding, as the highest values of $R_V$ are found in the regions with the highest levels of extinction (cluster \#11 and the supernebula).

In addition, \cite{plante_embedded_2002} show that a massive ($>10^6\,M_\odot$) star cluster is embedded in a dust cocoon that is composed of large dust grains, in contrast to what is found for the ISM dust. This is in line with the finding of larger values of $R_V$, which corresponds to larger grain size, at the location of the embedded SSCs here, as well as the scenario of cluster emergence sketched in the previous section.

\subsection{Electron temperature and density} \label{sec:discussion_Te}
\subsubsection{Comparison to literature}
In this section, we compare our results to measurements of electron temperature and density available from the literature.

MI10 presented a map of the [\ion{S}{II}]$\lambda$6731/6716 ratio based on FLAMES data. Our map shows good morphological agreement with theirs, although the FLAMES measurements, due to their lower spatial resolution, do not capture as much detail. The electron densities derived from our NFM data locally reach higher values compared to those reported by MI10 (i.e. a maximum of 1930\,cm$^{-3}$ as opposed to $\sim$800\,cm$^{-3}$ for the FLAMES data). This suggests that high-density structures are physically small with sizes $\lesssim17$\,pc. Indeed, our data reveals two high-density clumps that are barely resolved, with each clump measuring approximately 0\farcs1 in size and separated by about 0\farcs1.

\cite{guseva_vlt_2011} used UVES data to derive the electron temperatures and densities for four slit positions in NGC\,5253. The reader is referred to their Fig. 1 for the exact locations of the slits. At positions C1 and P2, which are located on the SSCs in the NE-SW direction and NW-SE direction (resp.), they determined $n_\mathrm{e}$([\ion{S}{II}]) to be $n_{e,\mathrm{C1}}=723\pm69$\,cm$^{-3}$ and $n_{e,\mathrm{P1}}=1108\pm99$\,cm$^{-3}$, similar to our findings. Morphologically, P2 stretches towards the North-Western bubble where we detect an enhancement in $n_\mathrm{e}$, consistent with the higher $n_\mathrm{e}$ for P2 than C1. Slit C2, located at the position of Complex \#2, they find $n_{e,\mathrm{C2}}=271\pm47$\,cm$^{-3}$, consistent with our measurements. Finally, at slit position P1, located across the top of the North-Western bubble, the authors find $n_{e,\mathrm{P1}}=366\pm50$\,cm$^{-3}$. This lies on the lower end of our measurements but remains within the same order of magnitude. 

\citet[hereafter MI12]{monreal-ibero_ionized_2012} determined the electron temperature in NGC\,5253 using FLAMES data. To determine the electron temperature for the low ionisation gas, they used the ratio [\ion{S}{II}]$\lambda$6717,6731/[\ion{S}{II}]$\lambda$4069,4076. They found that the temperature peaks in the proximity of the SSCs, reaching a maximum of $T_\mathrm{e}([\ion{S}{II}])=9700^{+1700}_{-1300}$\,K. This maximum temperature is lower than what we find in the NFM data (with $T_\mathrm{e,median}=11842 \pm 1667$\,K), but remains consistent within the uncertainties. 

For the temperature of the high-ionisation gas, we can also compare with MI12, although their analysis traces more highly ionised gas using the [\ion{O}{III}]$\lambda$(4959+5007)/4363 ratio. As in our results, the temperature increases near the SSCs, but this enhancement appears more pronounced in their data. Our temperature distribution is more uniform. MI12 report values ranging from approximately 9300 to 11000\,K, which is slightly lower than our average measurement from the NFM ($T_\mathrm{e,median}=11260\pm575$\,K). This could suggest that the high-temperature structures are confined to small spatial scales. Alternatively, the higher ionisation gas traced by O$^{++}$ has a lower $T_\mathrm{e}$ than that traced by S$^{++}$. 

The median value of the $T_\mathrm{e}$([\ion{S}{III}]) that we derive is consistent with the measurements of \cite{lopezsanchez_localized_2007}.

\subsubsection{Temperature - temperature relations} \label{sec:T-T relations}
The auroral lines used in the determination of electron temperatures are very faint so they often remain undetected. Additionally, the lines frequently fall outside of the available wavelength range of the spectrum. To still obtain estimates for the electron temperatures of different ionisation regimes (or zones), researchers have searched for relations between electron temperatures from different ions, using both observational studies \citep[e.g.][]{mendez-delgado_density_2023, rogers_chaos_2021, vaught_investigating_2024} and model-based approaches \citep[e.g.][]{garnett_electron_1992}. In this section, we assess how the temperatures derived in this work compare to these relations.

\begin{figure}
    \centering
    \includegraphics[width=\linewidth]{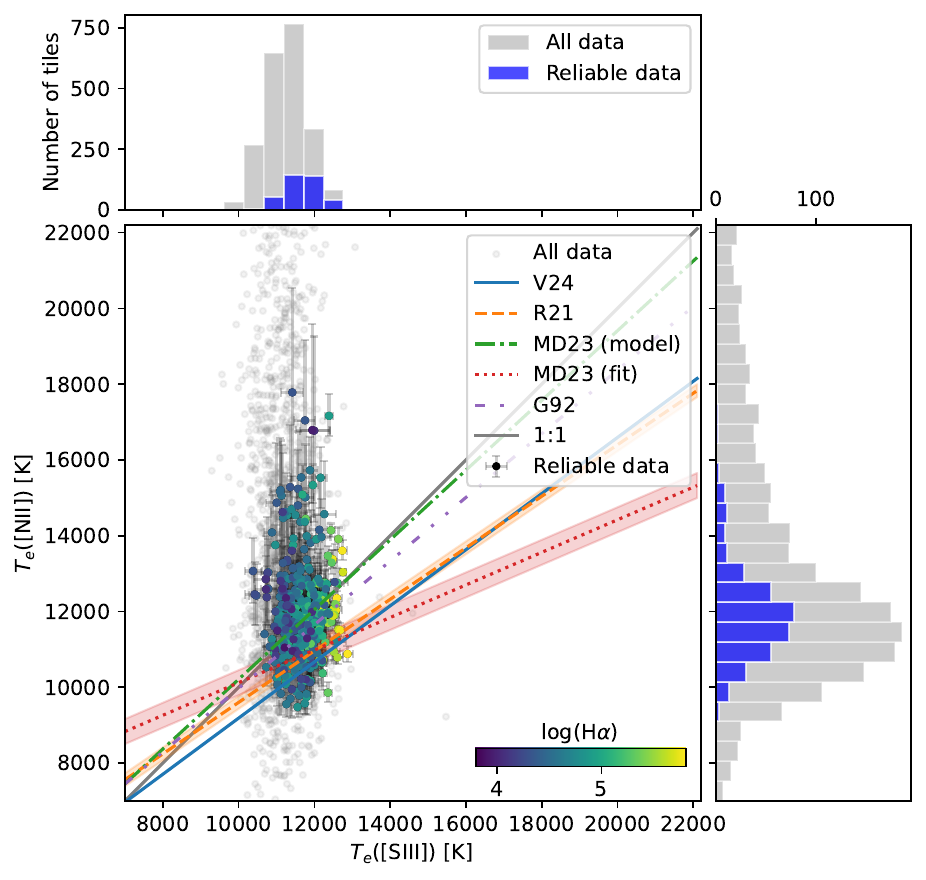}
\caption{$T_\mathrm{e}$([\protect\ion{S}{III}])-$T_\mathrm{e}$([\protect\ion{N}{II}]) relations from the literature: \protect\cite{vaught_investigating_2023} (V24; solid blue), \protect\cite{rogers_chaos_2021} (R21; dashed orange), \protect\cite{mendez-delgado_density_2023} (MD23) from the BOND models (dash-dot-dashed green) and observations (dotted red), \protect\cite{garnett_electron_1992} (G92; dash-dotted magenta), and the 1:1 line (solid grey), as well as their corresponding internal dispersion when known. The electron temperatures as determined for every Voronoi tile as determined in this work are plotted as points. As the reliable data, we use $T_\mathrm{e}$([\protect\ion{S}{III}]) determined by fixing $n_\mathrm{e}$([\protect\ion{Cl}{III}]) to its median value, and $T_\mathrm{e}$([\protect\ion{N}{II}]) only for those bins that have S/N([\protect\ion{N}{II}])5755$>$4. These measurements are colour-coded by the log of H$\alpha$ flux. The grey data points indicate the measurements for all tiles. The histograms show the number of tiles with a given $T_\mathrm{e}$ for all measurements (grey) and for the reliable measurements (blue).}
    \label{fig:TSIII-TNII}
\end{figure}

Figure \ref{fig:TSIII-TNII} shows how $T_\mathrm{e}$([\ion{N}{II}]) varies as a function of $T_\mathrm{e}([\ion{S}{III}])$ for each tile in our data. Literature relations are shown for comparison. We find that the range of determined temperatures is much larger for $T_\mathrm{e}$([\ion{N}{II}]) than for $T_\mathrm{e}$([\ion{S}{III}]). In general, $T_\mathrm{e}$([\ion{N}{II}]) is higher than expected from the published $T_\mathrm{e}$–$T_\mathrm{e}$ relations, with the highest S/N tiles clustering around the 1:1 line.

We find no strong evidence for a correlation between the two electron temperatures. This is supported by the Pearson correlation coefficient, $r = -0.025$, with a corresponding $p$-value of 0.6, which indicates no correlation. Instead, on average, the two temperatures, with reliable measurements, are similar, with around 40\% (160/393)  of reliable tiles having $0.95\leq T_\mathrm{e}$([\ion{S}{III}])/$T_\mathrm{e}$([\ion{N}{II}])$\leq1.05$ and 68\% (286/393) for a ratio between 0.9 $\leq T_\mathrm{e}$([\ion{S}{III}])/$T_\mathrm{e}$([\ion{N}{II}]) $\leq$ 1.1. 

These findings differ from most of the relations derived in the literature. This can potentially be explained in three different ways. Firstly, our data only covers a relatively small range in temperatures for $T_\mathrm{e}$([\ion{S}{III}]) in addition to a noisy measurement for $T_\mathrm{e}$([\ion{N}{II}]), preventing us from detecting a possible correlation. Then, an important difference between our work and the literature relations is that the measurements from literature are for a complete \ion{H}{II} region, while our points are measurements within a single tile and thus only encompass a small fraction of the \ion{H}{II} region. Therefore, this indicates that, possibly, relations derived for complete \ion{H}{II} regions cannot be applied to smaller scales.
Finally, our measurements appear consistent with the 1:1 line and the BOND models from \cite{mendez-delgado_density_2023}. These models give the temperature relations in the absence of temperature fluctuations. The agreement between our data and the BOND models may indicate that dividing the \ion{H}{II} region into small-scale tiles effectively removes the influence of temperature fluctuations -- something that affects the results when considering full \ion{H}{II} regions. However, an analysis of more \ion{H}{II} regions will be necessary to test this further.

\begin{figure}
    \centering
    \includegraphics[width=\linewidth]{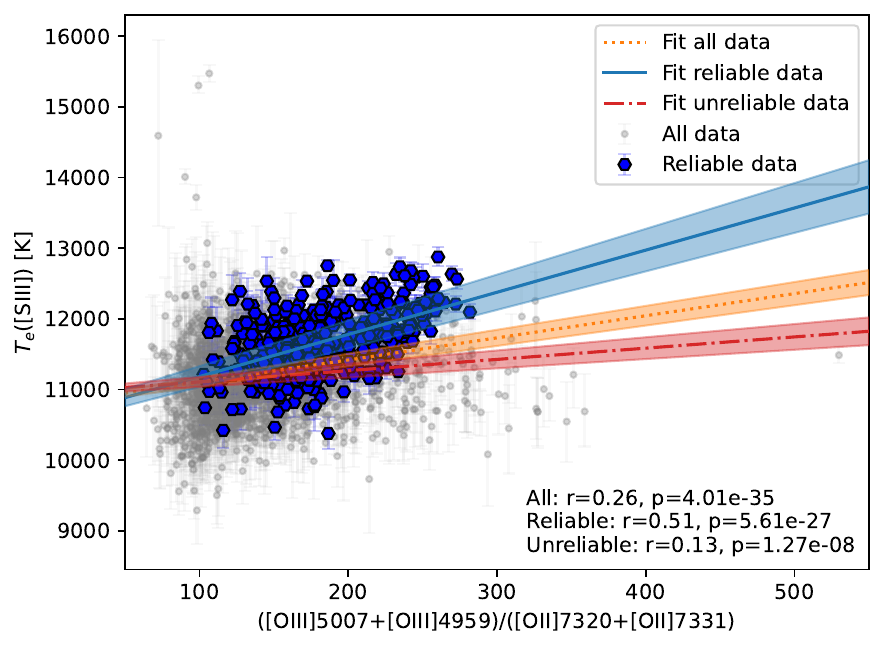}
    \caption{$T_\mathrm{e}$([\ion{S}{III}]) as a function of level of ionisation (([\ion{O}{III}]5007+[\ion{O}{III}]4959)/([\ion{O}{II}]7320+[\ion{O}{II}]7331)). The measurements from all tiles are plotted in grey, the dark blue hexagons indicate the reliable data (S/N([\ion{N}{II}])5755$>$4). The solid blue line indicates a linear fit through the reliable data, the orange dotted line through all data, and the red dash-dotted line only through the data points with S/N([\ion{N}{II}])5755$<$4. The text in the lower-right corner indicates the Pearson coefficient and the p-value of the different groups of data.}
    \label{fig:OIII_OII_TeSIII}
\end{figure}

In Fig. \ref{fig:TSIII-TNII}, it can be seen that $T_\mathrm{e}$([\ion{S}{III}]) is correlated with the H$\alpha$ flux, with points with a higher H$\alpha$ flux having higher electron temperatures. To first order, the H$\alpha$ flux is a proxy for the distance from the SSCs, but better accounts for the complex morphology than e.g. a projected distance would. We can therefore say that $T_\mathrm{e}$([\ion{S}{III}]) decreases when we move away from the SSCs. Another way to examine this is by plotting $T_\mathrm{e}$([\ion{S}{III}]) as a function of the level of ionisation, in this case traced by ([\ion{O}{III}]5007+[\ion{O}{III}]4959)/([\ion{O}{II}]7320+[\ion{O}{II}]7331), as is shown in Figure \ref{fig:OIII_OII_TeSIII}. Indeed, a strong correlation between $T_\mathrm{e}$([\ion{S}{III}]) and the level of ionisation is visible. The correlation becomes stronger when we only look at the points with S/N([\ion{N}{II}])5755$>$4, which are also the points closest to the SSCs. From this, we can conclude that the gas traced by S$^{++}$ is sitting in the 3D vicinity of the SSCs and tied directly to their presence.
In contrast, no such clear conclusions can be drawn for the gas traced by N$^+$, because (1) $T_\mathrm{e}$([\ion{N}{II}]) does not show a relation with  $T_\mathrm{e}$([\ion{S}{III}]) and (2) $T_\mathrm{e}$([\ion{N}{II}]) does not show a correlation with ([\ion{O}{III}]5007+[\ion{O}{III}]4959)/([\ion{O}{II}]7320+[\ion{O}{II}]7331) (plot not shown). This is primarily due to the large scatter in $T_\mathrm{e}$([\ion{N}{II}]), caused by the large uncertainties. Thus, it is difficult to say if the lack of correlation is physical, i.e. the gas traced by N$^+$ is located in front of the SSCs and not directly related, or if there is a correlation hidden in the uncertainties. Nevertheless, the increase in S/N of the N$^+$ line in the proximity of the SSCs suggests a connection.

\subsection{Possible sources of nitrogen enrichment} \label{sec:discussion_NO}
In this section, we will discuss the possible origins of the observed nitrogen enrichment (see Fig. \ref{fig:logNO}). We will also put our observations in the context of results at high redshift.

\subsubsection{Wolf-Rayet stars} \label{sec:discussion_N_WRstars}
The prime candidate for the source of nitrogen enrichment in NGC\,5253 are WR stars \citep[][among others]{walsh_optical_1989, kobulnicky_hubble_1997,lopezsanchez_localized_2007}, even though some parts of the galaxy appear chemically uniform despite the proximity of WR stars \citep{monreal-ibero_study_2010,westmoquette_piecing_2013}. In this section, we discuss the relation between the nitrogen enrichment and WR stars in the central area of NGC\,5253. We note that from the WFM observations alone it is impossible to distinguish between the signatures of very massive stars (VMS, $M\gtrapprox100\odot$) and WR stars, since the discriminatory features at the position \citep{martins_inferring_2023} of the blue WR bump are outside our available wavelength range. Therefore, for this discussion we assume all detected emission comes from WR stars. However, the presence of VMS has been detected in cluster \#5 \citep{smith_very_2016}.

\begin{figure}
    \centering
    \includegraphics[width=\linewidth]{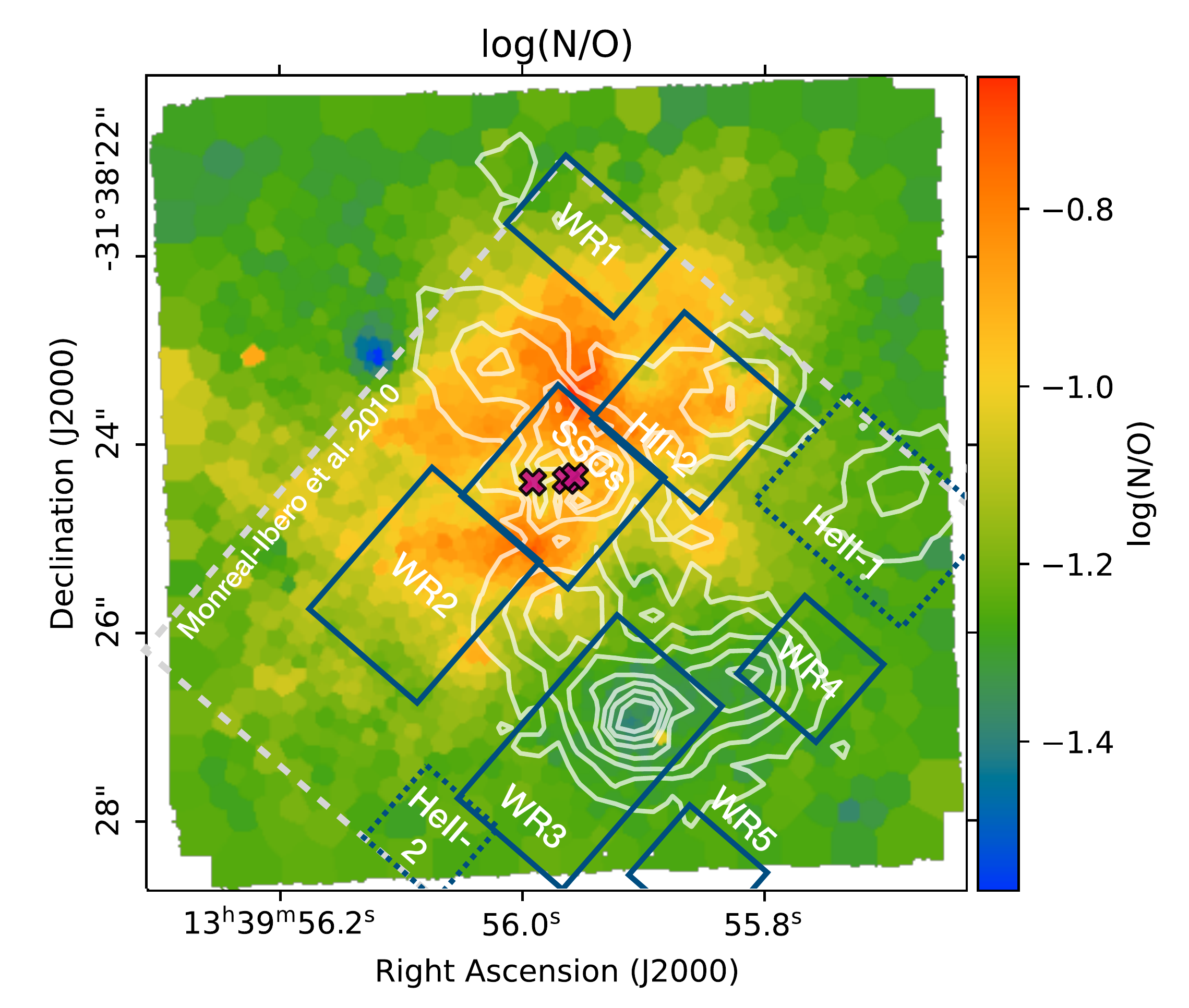}
    \caption{Map of N/O abundance overlaid with red WR bump luminosity contours (white) and the position of the blue bump by MI10 (dark blue rectangles). The magenta crosses indicate the positions of the SSCs.}
    \label{fig:NO_WR}
\end{figure}

We compare the spatial distribution of WR stars and nitrogen enrichment in Fig. \ref{fig:NO_WR}. We find no clear spatial overlap. The peak in N/O is located between the northern group of WR stars, while Complex \#2 -- where the highest concentration of WR stars is found -- shows no excess in nitrogen. Although some nitrogen enhancement is observed at the location of the SSCs, it is less pronounced than in other regions. This finding aligns with \cite{james_lyman_2013} for the BCD galaxy Haro 11, who also report a low N/O ratio in regions with many WR stars and a high N/O in regions with fewer. They hypothesise that in areas with numerous WR stars, the N-rich material has not yet mixed with the warm ionised gas, whereas in regions with few WR stars, the stars have likely completed their WR phase, and the expelled nitrogen has had time to mix. 

The first question we address is whether the existing WR stars could have produced enough nitrogen to explain the observed enhancement. Following an approach similar to \cite{brinchmann_galaxies_2008}, we calculate the mass of the additional nitrogen via

\begin{equation} \label{eq:MN}
    \Delta M_N = \frac{m_N}{m_H} M_{\text{ionised}} \Delta\left(\frac{N}{H}\right)
\end{equation}

\noindent where $m_N$ and $m_H$ are the mass of a nitrogen and hydrogen atom, $\Delta\left(\frac{N}{H}\right)$ is the surplus abundance of nitrogen w.r.t. the median, and $M_{\text{ionised}}$ is the total mass of the ionised hydrogen obtained via equation \ref{eq:Mionised} \citep[e.g.][]{dopita_astrophysics_2003}.

\begin{equation}\label{eq:Mionised}
    M_{\text{ionised}} = \frac{m_H L_{H\beta}}{1.235\times 10^{-25}T_4^{-0.86}n_\mathrm{e}}
\end{equation}

\noindent where $L_{H\beta}$ is the (extinction corrected) H$\beta$ luminosity, $T_4$ is the T$_e$([\ion{N}{II}]) in units of 10$^4$\,K, and $n_\mathrm{e}$ the [\ion{S}{II}] electron density. We estimate the mass of the extra nitrogen to be $\Delta M_{\text{N,NGC5253}} \approx 0.3\,M_\odot$. This is lower than the estimate of 1-10\,$M_\odot$ by \cite{smith_very_2016}, this is because they assume a lower N/O abundance for the unenriched gas: log(N/O)=-1.37 instead of log(N/O)=-1.12 in combination with a uniform cylindrical distribution of the nitrogen enhancement.

We can compare the mass of extra nitrogen to the estimated mass in nitrogen yield from the observed WR stars. Using Eq. \ref{eq:MN} again, we now take $N/H$ as the nitrogen abundance in the WR star shell and $M_{\text{ionised}}$ as the total mass lost by the star. We adopt a typical mass-loss rate of $10^{-5}\,M_\odot\,\text{yr}^{-1}$ over 2\,Myr \citep{crowther_physical_2007,brinchmann_galaxies_2008}, yielding a total of $\sim20\,M_\odot$. Assuming that the ring nebula NGC\,6888 \citep{esteban_chemodynamics_1992} is representative of a typical WR star and taking the nitrogen abundance as 12+log(N/H)=8.1 from the shell, we estimate that the observed WN stars should be able to produce $\sim1.7\,M_\odot$ of nitrogen. 

Alternatively, we can make the calculation using more recent measurements of MI-76 \citep{fernandez-martin_integral_2013}. MI-76 has an estimated mass loss of $10^{-3}\,M_\odot\,\text{yr}^{-1}$ over $10^4$ years (i.e. 10\,$M_\odot$ total) and a nitrogen abundance of 12+log(N/H)=7.92 (in region 7). From this, we estimate a total mass loss in nitrogen of $M_{\text{N,MI-76}}\approx0.01\,M_\odot$. This estimate is a lower limit, because the region 7 has the lowest nitrogen abundance. Assuming this amount of mass loss applies to all WN stars in NGC\,5253, this accounts for a total nitrogen yield of 0.3\,$M_\odot$. 

Thus, we conclude that the WR stars are easily capable of producing the observed nitrogen overabundance of 0.3\,$M_\odot$.

If WR stars are responsible for the nitrogen enrichment, one would also expect an accompanying enhancement in helium abundance \citep[e.g.][]{schaerer_about_1996, kobulnicky_hubble_1997}. We do not detect such an enhancement in helium abundance in our data. However, this He enhancement is at a much lower level than the enhancement in nitrogen \citep[e.g.][]{brinchmann_galaxies_2008}. Moreover, the original He abundance is much higher than that of N, making small enhancements proportionally harder to detect. \cite{monreal-ibero_he_2013} were unable to find conclusive evidence for enhancement in helium in the FLAMES dataset. Deeper data is therefore necessary to draw reliable conclusions on this.

Having established that the currently observed WR stars are capable of producing sufficient nitrogen, we now assess whether this nitrogen could have travelled to the locations where enrichment is detected. It seems plausible that the increase in N/O directly to the north of the SSCs is caused by the expulsion of N-rich gas by the clusters directly surrounding this location, i.e. WR1, HII-2, SSCs, but potentially also the WN stars in regions H, B, A and C before they evolved into WC stars. 

As an upper limit, we consider the much larger distance between the peak of nitrogen enrichment and Complex \#2 -- containing the largest number of WR stars -- which is approximately 60\,pc. If we assume the lifetime of a WR star to be $\sim1$\,Myr \citep{crowther_physical_2007}, a velocity of the nitrogen-enriched material of 60\,km/s is sufficient to cover this distance. This is orders of magnitude slower than the velocity of the winds expelled by the stars of $700\leq v_\infty\,[\mathrm{km/s}] \leq 2200$ \citep{crowther_physical_2007}, and of the same order as the maximum radial velocity of [\ion{N}{II}]$\lambda$6584. Nevertheless, this is the maximum distance the N-rich material has had to travel. In a more likely scenario, the N-rich material observed at the peak of enrichment comes from the directly surrounding WR stars, as discussed above.

This suggests a scenario in which the WR stars in Complex \#2 uniformly expel their nitrogen-rich gas. However, the observed nitrogen enhancement is only observed towards the north of Complex \#2. We propose that the N-rich material moving in the southern direction gets dispersed and diluted too much for us to be able to detect, while the large increase in electron density around the SSCs (see Fig. \ref{fig:Tene_low_Vorbin}) impedes gas flow, causing it to stall and mix locally with the ISM. In fact, the positions showing high N/O seem to concentrate around regions with a high density, supporting this theory. Combined with the presence of surrounding WR stars, this may also explain the plume-like structure of the enriched region, which appears to be funnelled between two high-density areas. 

Alternatively, the lack of N-rich gas at the position of Complex \#2 can be explained in a scenario similar to \cite{james_lyman_2013} in 
Haro 11 and also proposed by \cite{monreal-ibero_ionized_2012} for NGC\,5253, where they suggest there has not been sufficient time between the WN stars expelling the N-rich material and for this material to subsequently mix with the warm ionised gas. At this stage, the uncertainties in N-production per WR star and the estimations of the number of WR stars are too large to make a distinction between this scenario and the one described above. Perhaps with even higher spatial resolution data, one should be able to observe small-scale N-enhancement at the position of (individual) WR stars.

\subsubsection{Other sources of nitrogen enrichment}
In this section we will discuss other proposed candidate enrichers and how likely they are to have caused the observed enrichment in NGC\,5253. The candidates are supernova (remnants), asymptotic giant branch (AGB) stars, and very massive stars (VMS).

\cite{smith_very_2016} identified the presence of VMS in cluster \#5. The authors reason that the stellar winds of massive ($>50\,M_\odot$) rotating cluster O stars in clusters \#5 and \#11 are likely capable of producing $>1\,M_\odot$ N-rich gas in the first 1\,Myr of their lives. This is in principle sufficient to explain the observed mass in extra nitrogen. If this were the source of the observed N abundance, the N-rich gas must have moved away from the SSCs, because we only detect a marginal amount of enrichment at that position. Therefore, VMS are an additional explanation of the enrichment. As discussed in Sect. \ref{sec:discussion_N_WRstars} with the optical spectra provided by MUSE alone it is impossible to distinguish the spectra from VMS from that of the WR stars, and we have treated them as the same objects in this work. 

Two supernovae have been detected in NGC\,5253: SN\,1895B \citep{pickering_new_1895} and SN\,1972E. However, both SN are located far away from the area of observed nitrogen enrichment and are therefore unlikely to be its origin. 
Out of the seven supernova remnant (SNR) candidates identified through the near-infrared [\ion{Fe}{II}] emission line signature by \cite{labrie_near-infrared_2006}, only N5253-S007 is located in the proximity of the nitrogen enrichment. Nevertheless, the [\ion{Fe}{II}] ratio is low and reliable measurements are complex due to the strong emission coming from its direct surroundings. No evidence is found for synchrotron emission from SNR and the emission coming from the SSCs is completely thermal \citep{beck_central_1996, turner_radio_2000}. This means that supernova (remnants) are unlikely to be the origin of the nitrogen enrichment.

AGB stars have atmospheres enriched with nitrogen. They are stars in a late stage of stellar evolution, taking about 100\,Myr-10\,Gyr to turn off the main sequence \citep{karakas_stellar_2007}. This is much older than the ages of the cluster surrounding the area of nitrogen enrichment \citep{de_grijs_ngc_2013}. While clusters of this age have been detected in this galaxy, they are located far away from where the nitrogen overabundance is observed and outside the FOV of the NFM. Therefore, we can rule out AGB stars as the source of nitrogen enrichment.

\subsubsection{Comparison to high-redshift}
\begin{figure}
    \centering
    \includegraphics[width=\linewidth]{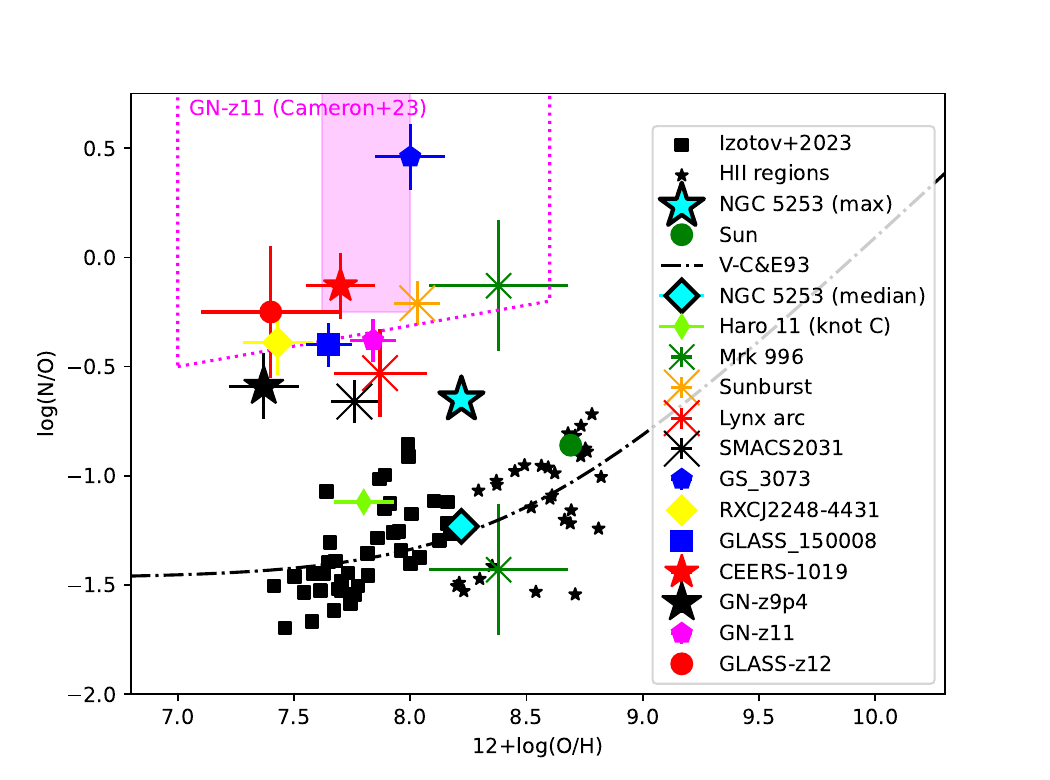}
    \caption{Figure adapted from \protect\cite{marques-chaves_extreme_2024}. The cyan star with black border indicated the most extreme value observed in NGC\,5253 from this work, while the cyan diamond with a black border represents the median value for the NFM FOV. Local galaxies \protect\citep{izotov_abundances_2023} and \protect\ion{H}{II}-regions \protect\citep{esteban_optical_2002,esteban_keck_2009,esteban_carbon_2014} are shown in black. The magenta region is the range of allowed abundances for GN-z11 \protect\citep{cameron_nitrogen_2023}. The red star indicates CEERS-1019 \protect\citep{marques-chaves_extreme_2024}. The green crosses indicate two values for the BCD Mrk996: the upper point for the enriched central region and the lower for the galaxy average \protect\citep{james_vlt_2009}. The red cross marks the (gravitationally lensed) Lynx arc \protect\citep{villar-martin_nebular_2004}; the black cross SMACS2031 \protect\citep{patricio_young_2016}; the black star GN-z9p4 \protect\citep{schaerer_discovery_2024}; the yellow diamond RXCJ2248 \protect\citep{topping_metal-poor_2024}; the blue square GLASS-z12 \protect\citep{castellano_jwst_2024}; the blue pentagon GS\_3037 \protect\citep{ji_ga-nifs_2024}; the neon green narrow diamond is Knot C in the BCD Haro 11 \protect\citep{james_lyman_2013}; and the orange cross the (gravitationally lensed) sunburst arc \protect\citep{pascale_nitrogen-enriched_2023}. The dash-dotted line indicates the relation for low-z star-forming galaxies by \protect\cite{vila-costas_nitrogen--oxygen_1993}.}
    \label{fig:compare_highz}
\end{figure}

We compare our results with chemical enrichment trends observed both locally and at high redshift, as shown in Fig. \ref{fig:compare_highz}. The most extreme region in NGC\,5253 exhibits a metallicity of 12+log(O/H) = 8.2 and log(N/O) = –0.65. This places it significantly above typical local galaxies and \ion{H}{II} regions in terms of nitrogen enrichment \citep{izotov_abundances_2023}, and in a similar regime as the N-rich high-redshift galaxies, whereas the median value across our field of view in NGC\,5253 aligns well with the properties of local galaxies.

Recent studies propose AGN \citep{isobe_jades_2025} and globular clusters with multiple populations \citep{Charbonnel2023,DAntona2023} as possible origins of the nitrogen enhancement. While the N-rich material in NGC\,5253 is concentrated around the SSCs, very little enrichment is observed at the exact location of these clusters. This poses a challenge for scenarios in which the GCs have to hold on to their N-rich material.
In addition, the lack of a (intermediate mass) black hole (BH) in NGC\,5253 showcases that nitrogen enrichment is possible without the presence of an AGN.

WR features have been detected for the Sunburst Arc \citep[z=2.396][]{RiveraThorsen2024}, the BCD galaxy Mrk\,996 \citep{james_vlt_2009}, and the BCD Haro\,11 \citep{james_lyman_2013}. For the highest redshift sources ($z>3$), there are no WR features detected. However, this does not mean that they might not be present, because the features are weak and therefore might fall below the detection threshold.

The detailed study of BCDs is valuable for the understanding of nitrogen enrichment in high-z galaxies, especially in light of the association of nitrogen-rich galaxies with small sizes \citep{harikane_jwst_2025}. One caveat is that we did not consider how these effects would change when accounting for observational effects when observing high-z galaxies. For example, luminosity weighting is expected to have a large impact on the observations. To fully understand how well the analogy between this high-resolution study and observations of high-z galaxies holds, studies simulating these observations at high redshift are desirable. In addition, the simulation of local observations at high redshift presents itself as a nice preparatory science for research done with future facilities, like HARMONI and MOSAIC at the Extremely Large Telescope. 

\section{Conclusions} \label{sec:conclusions}
We presented a thorough study of the central 8\farcs1 × 7\farcs5 of the BCD galaxy NGC\,5253, observed with MUSE at the VLT with the NFM configuration. We leverage our high spatial resolution (0\farcs15 $\sim$ 2.3\,pc) to study the gas and dust surrounding the three SSCs at unprecedented detail. The main results and conclusions are presented here.

\begin{itemize}
    \item We present here, to our knowledge, the first ever extragalactic spatially resolved mapping of the attenuation law through means of $R_V$. We show that the shape of the attenuation law is variable at the scale of individual clusters. We find the embedded supernebula has the highest $R_V$ ($R_V=9.7 \pm 1.5$), indicative of larger dust grains, while clusters \#5 and \#11 show lower but still elevated values compared to the MW average ($R_V=3.9 \pm 0.5$ and $R_V=6.9 \pm 0.7$). These findings support an evolutionary sequence: the supernebula represents a cluster that is still deeply embedded in its natal gas cloud, while clusters \#5 and \#11 are more evolved, having emerged from their natal clouds, with \#5 in a more advanced stage of emergence. 

    \item The low-ionisation plasma shows a uniform electron temperature distribution, with a median of $T_\mathrm{e}$([\ion{N}{II}])$=11800\pm1700$\,K, but a structured density map with a maximum value of $n_\mathrm{e}=1930\pm40$\,cm$^{-3}$ found slightly south of the position of cluster \#5 and second maximum of $n_\mathrm{e}=1800\pm50$\,cm$^{-3}$ 0\farcs1 to the south of this first maximum. The high-ionisation plasma has a flat temperature distribution, with a median of $T_\mathrm{e}$([\ion{S}{III}])$=11250\pm575$\,K.

    \item We find a nearly uniform distribution of helium with a mean abundance ($\pm$ standard deviation) of $10^3y^+=81 \pm 4$. We identify two regions encompassing a cluster with enhanced helium abundance, of which one of them is suggested to be associated with a WR population.

    \item We find a uniformly distributed total oxygen abundance of 12+log(O/H)=$8.22\pm0.05$.

    \item For the first time, we map the nitrogen enrichment at 2.3 pc resolution. We clearly detect a factor 2-3 nitrogen enrichment compared to the surrounding ISM. The peak of enrichment is located north of the SSCs and no enrichment is detected at the position of Complex \#2. While the N-rich material is concentrated around the SSCs, very little is found at the exact position of the clusters.

    \item We map the WC-type Wolf-Rayet star population via the red bump using the MUSE WFM data. We estimate the number of WC stars to be between 12 and 20. In addition, we estimate the number of WN-type stars from the blue bump detected by MI10 in the FLAMES data to be around 21-25. This means that we detect a total of $\sim40$ WR stars. We estimate that the observed WR stars are able to produce enough nitrogen to explain the observed extra mass in nitrogen of 0.3\,$M_\odot$. Because there is no direct spatial overlap between the enrichment and WR star positions, the N-rich material appears to have been expelled from the original sites.
    \item We show that the region of most extreme enrichment in NGC\,5253 has values comparable to high-redshift galaxies. The median value of the central area aligns well with local galaxies. 
    
\end{itemize}

\section*{Acknowledgements}
P.M.W. gratefully acknowledges support by the BMBF from the ErUM program (project BlueMUSE-CD, grant 05A23BAC).
L.A.B. acknowledges support from the Dutch Research Council (NWO) under grant VI.Veni.242.055 (\url{https://doi.org/10.61686/LAJVP77714}) and the ERC Consolidator grant 101088676 ("VOYAJ”).
We are also grateful to the communities who developed the many Python packages used in this research, such MPDAF \citep{Piqueras2017}, Astropy \citep{astropy2013}, numpy \citep{Walt2011}, scipy \citep{Virtanen2020}, and matplotlib \citep{Hunter2007}.

\section*{Data Availability}
This project uses data obtained with the Multi Unit Spectroscopic
Explorer (MUSE) under proposal ID 104.A-0026
(PI: L. WISOTZKI) for the Narrow Field Mode. The MUSE Wide Field Mode data is obtained under proposal ID 095.B-0321 (PI: L. VANZI). All the raw and pipeline processed data used in this study are publicly available from the ESO Science Archive \url{https://archive.eso.org/scienceportal/home}.




\bibliographystyle{mnras}
\bibliography{references_abbreviation_080925,references2} 




\appendix
\section{Atomic data} \label{sec:appendix_atomicdata}
Table \ref{tab:AtomicData} gives the references to the atomic data used in \texttt{PyNeb} for this paper. This is for version \texttt{PYNEB\_23\_01}.
\begin{table*}
    \centering
    \begin{tabular}{c c c c}
        \thickhline
        Ion & $A$-levels & Energy levels & Collisional strengths \\
        \thickhline
        O$^0$ &\cite{wiese_atomic_1996} & \cite{7288EL} & \cite{bhatia_neutral_1995}\\
        O$^+$ & \cite{wiese_atomic_1996} & \cite{10.1063/1.555928} & \cite{kisielius_electron-impact_2009}\\
        O$^{++}$ & \cite{storey_theoretical_2000}; \cite{fischer_new_2004} & \cite{7288EL} & \cite{storey_collision_2014} \\
        S$^+$ & \cite{rynkun_theoretical_2019} & \cite{10.1063/1.555862} & \cite{tayal_breit-pauli_2010} \\
        S$^{++}$ & \cite{FROESEFISCHER2006607}  & \cite{10.1063/1.555862} & \cite{tayal_collision_1999} \\
        \thickhline
        H & \cite{storey_recombination_1995} & \cite{KRAMIDA2010586} & -\\
        He & \cite{porter_improved_2012}; \cite{porter_erratum_2013} & - & - \\
        \thickhline
    \end{tabular}
    \caption{Default atomic data used in \texttt{PyNeb}. The left-most column indicates for which ion the data is listed. For the recombination atoms we list the atom instead. The second column gives the Einstein $A$ coefficient. The third column gives the energy level data. Finally, the last column lists the references for the collisional strengths.}
    \label{tab:AtomicData}
\end{table*}

\section{Ionic oxygen abundances}
Here we present the maps of the determined ionic oxygen abundances as described in Sect. \ref{sec:results_abundances}. 
\begin{figure*}
    \centering
    \includegraphics[width=\linewidth]{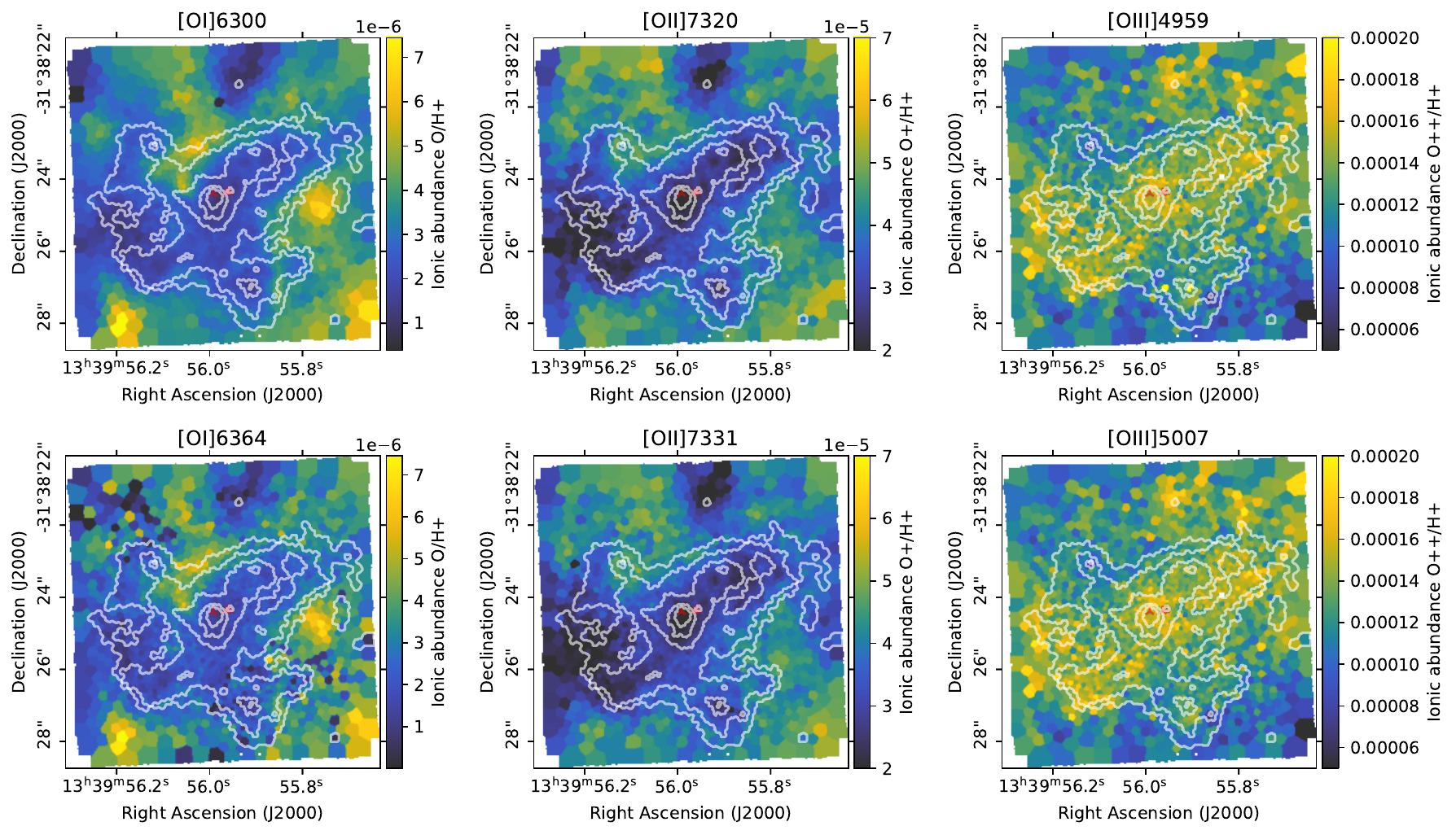}
    \caption{Ionic oxygen abundance for each of the collisional excited transitions displayed in the caption of each panel. From left to right, we show abundances for O, O$^+$, and O$^{++}$. Top and bottom show two transitions observed of the same ion. The colourscale for the same ion is defined to be equal.}
    \label{fig:IonicOxygen}
\end{figure*}


\bsp	
\label{lastpage}
\end{document}